\documentclass[review]{elsarticle}

\usepackage{amsmath}
\usepackage{lineno,hyperref}
\modulolinenumbers[5]
\usepackage{caption}
\usepackage{subcaption}
\usepackage{nicefrac}
\usepackage{booktabs}
\usepackage[usenames]{xcolor}
\usepackage{units}
\usepackage[acronyms]{glossaries}

\journal{Combustion and Flame}

\makeatletter
\providecommand{\doi}[1]{%
  \begingroup
  \let\bibinfo\@secondoftwo
  \urlstyle{rm}%
  \href{http://dx.doi.org/#1}{%
    doi:\discretionary{}{}{}%
    \nolinkurl{#1}%
  }%
  \endgroup
}
\makeatother

\newcommand{\ud}{\,\mathrm{d}}

\newcommand{\wt}[1]{\widetilde{#1}}
\newcommand{\ol}[1]{\overline{#1}}

\newacronym{dns}{DNS}{direct numerical simulation}
\newacronym{les}{LES}{large eddy simulations}
\newacronym{pdf}{PDF}{probability density function}
\newacronym{pmf}{PMF}{probability mass function}
\newacronym{fpdf}{FPDF}{filtered probability density function}
\newacronym{fdf}{FDF}{filtered density function}
\newacronym{ml}{ML}{machine learning}
\newacronym{dnn}{DNN}{deep neural network}
\newacronym{vae}{VAE}{variational auto-encoder}
\newacronym{cvae}{CVAE}{conditional variational autoencoder}
\newacronym{gan}{GAN}{generative adversarial network}
\newacronym{ols}{OLS}{ordinary least squares regression}
\newacronym{fgm}{FGM}{flamelet-generated manifolds}
\newacronym{dof}{DoFs}{degrees of freedom}
\newacronym{rmse}{RMSE}{root mean square error}
\makeglossaries
\glsdisablehyper

\begin{document}

\begin{frontmatter}

\title{Deep learning for presumed probability density function models}

\author[main]{M.~T.~Henry de Frahan\corref{cor1}}
\ead{marc.henrydefrahan@nrel.gov}

\author[main]{S.~Yellapantula}

\author[main2]{R.~King}

\author[lbnl]{M.~S.~Day}

\author[main]{R.~W.~Grout}
\ead{ray.grout@nrel.gov}

\cortext[cor1]{Corresponding author}

\address[main]{High Performance Algorithms and Complex Fluids, Computational Science Center, National Renewable Energy Laboratory, 15013 Denver W Pkwy, ESIF301, Golden, CO 80401, USA}
\address[main2]{Complex Systems Simulation and Optimization Group, Computational Science Center, National Renewable Energy Laboratory, 15013 Denver W Pkwy, ESIF301, Golden, CO 80401, USA}
\address[lbnl]{Center for Computational Sciences and Engineering, Lawrence Berkeley National Laboratory, Berkeley, CA 94720, USA}

\begin{abstract}
  In this work, we use \gls{ml} techniques to develop presumed
  \gls{pdf} models for \acrlong{les} of reacting flows. The joint
  sub-filter \gls{pdf} of mixture fraction and progress variable is
  modeled using various \gls{ml} algorithms and commonly used
  analytical models. The \gls{ml} algorithms evaluated in the work are
  representative of three major classes of \gls{ml} techniques:
  traditional ensemble methods (random forests), deep learning
  (\acrlong{dnn}s), and generative learning (\gls{cvae}). The first
  two algorithms are supervised learning algorithms, and the third is
  an unsupervised learning algorithm. Data from \acrlong{dns} of the
  low-swirl burner~\cite{Day2012} are used to develop training data
  for sub-filter \gls{pdf} models. Models are evaluated on predictions
  of the sub-filter \glspl{pdf} as well as predictions of the filtered
  reaction rate of the progress variable, computed through an integral
  of the product of the sub-filter \gls{pdf} and the conditional means
  of the reaction rate. This a-priori modeling study demonstrates that
  deep learning models for presumed \gls{pdf} modeling are three times
  more accurate than analytical $\beta$-$\beta$ \gls{pdf} models and
  linear regression models. These models are as accurate as random
  forest models while using five times fewer trainable parameters and
  being 25 times faster for inference. In this work, conditional
  unsupervised learning did not present additional advantages beyond
  supervised learning with a feed-forward neural network. We
  illustrate how models generalize to other regions of the flow and
  develop criteria based on the Jensen-Shannon divergence to quantify
  the performance of a model on new data.
\end{abstract}


\end{frontmatter}

\glsresetall


\section{Introduction}
Simulation has the potential to accelerate the development of
cost-effective combustion technologies. Even with modern
high-performance computing hardware however, the computational cost of
fully resolving the reacting flows in these devices can be
prohibitive. \Gls{les} reduce the computational burden of simulating
turbulent reacting flows. \Gls{les} work with spatially filtered state
variables, which exhibit considerably less temporal and spatial
structure and thus require much less numerical resolution. However,
physical processes occurring at scales smaller than the filter width
must then be approximated with ``closure models,'' which of course
then determine the accuracy of the approach. \Gls{les} closure models
for nonreacting flows have received a great deal of recent attention,
and they are now in standard use for a wide range of engineering
applications. For reacting flows, considerable complexity arises from
the necessity to incorporate additional fine scales because of
chemical processes and chemistry-turbulence interactions. One approach
to constructing sub-filter \gls{les} models for reacting flows is to
express modeled quantities as weighted integrals between the physical
state and a \gls{pdf}. A presumed \gls{pdf} approach posits a class of
parameterized functional shapes for such \glspl{pdf}, and thus it
defines a parameterized model based on the resulting weighted
integrals. In some of the earliest work in this area, \citet{Cook1994}
proposed the use of $\beta$ functions for the \gls{pdf} shape for a
conserved scalar such as mixture fraction; much of the work in the
field since then has followed this basic strategy. \citet{Jimenez1997}
provided further analysis to justify the appropriateness of the
$\beta$ \gls{pdf} for passive scalar mixing. \citet{Bradley1998,
  Bradley2002} investigated a mixedness-reactedness formalism to
increase model fidelity. \citet{Ihme2008, Ihme2008a} determined that
the ``statistically most likely distribution'' was most appropriate
for a reacting scalar case.

The objective of the work presented here is to expand on the presumed
$\beta$ approach, with specific focus on the case of
reacting scalars. We incorporate a variety of \gls{ml} algorithms to
explore the accuracy of a number of \gls{pdf} shape functionals for
their use with an \gls{les} model, and we judge them by their ability to
reproduce a large-scale, \gls{dns} data set for a specific reacting flow
configuration. We explore three major classes of \gls{ml} algorithms
for use in this context: traditional ensemble methods (random
forests); deep learning (deep neural network); and deep, generative,
unsupervised learning (conditional variable autoencoder). More
broadly, traditional \gls{ml} methods include techniques such as
linear and polynomial regression, k-nearest neighbors, support vector
machines, Gaussian processes, and random forests. Of these, we focus
only on the latter because they have demonstrated widespread success for
complex modeling
applications~\cite{Fernandez-Delgado2014,Liaw2002}. Random forests are
based on an ensemble of decision trees, where decisions are based on
the model parameters to provide estimates of the
target. \Glspl{dnn} are universal function
approximators~\cite{Cybenko1989,Hornik1991} based on a sequence of
learnable linear operators and activation functions that are tuned
using a gradient-descent optimizer. \Glspl{dnn} have received much
attention in recent years, in large part because of the availability of large
public training data sets and powerful computing platforms such as
graphics processing units (GPUs)~\cite{Goodfellow2016}. Additional
advances in deep learning, particularly in the types of neural
network architectures, have led to breakthroughs in generative and
unsupervised learning, where new data are generated using the models
with unlabeled data and then by identifying trends and commonalities in
the generated data. \Glspl{vae} leverage neural networks to encode
information from input data into a latent space, which can then be
sampled through a decoder to generate new distributions that are
similar to the original data set.

In this work, we use the three \gls{ml} approaches discussed here
to construct presumed \gls{pdf} models for a \gls{dns} data set
that is a snapshot of a quasi-stationary simulation of a low-swirl,
premixed methane-air burner~\cite{Day2012}. We then evaluate the
suitability of the different classes of \gls{ml} algorithms, and of
the presumed \gls{pdf} model itself, both for data from a subregion of
the \gls{dns} and for the entire simulation
domain. In Section\,\ref{sec:formulation}, we formulate the target
problem and methods, including the details of the presumed \gls{pdf}
approach, the \gls{dns} target data, and the \gls{ml} algorithms and
network architectures explored. In Section\,\ref{sec:results}, we
compare the \gls{ml}-based constructions to simple analytic models.

\section{Formulation}\label{sec:formulation}

\subsection{Presumed \acrlong{pdf} modeling for combustion}\label{sec:physics}

In \gls{les} of reacting flows using presumed forms of \glspl{pdf}, an
important unclosed term in the equations is the filtered reaction
rates, appearing as a source term in the transport equation for
species mass fractions or progress
variables~\cite{Veynante2002,Pitsch2006a}. A common approach to
modeling the filtered reaction rates is to express it as an integral
of the product of a reaction rate derived from a physical model and a \gls{pdf}. The
conditioning variables are typically chosen to correlate strongly with
mixing (mixture fraction) and flame propagation (progress variable)
space~\cite{Bradley1998, Bradley2002}, accounting for much of the subgrid variation about the
mean. The conditional rate can then be modeled through a variety of
approaches to identify the manifold, such as canonical calculations
and tabulation (e.g., \acrlong{fgm}~\cite{VanOijen2002}, flame
prolongation of intrinsic low dimensional
manifold~\cite{Gicquel2000}), solving conditional transport equations
(e.g., conditional moment closure~\cite{Klimenko1999}), or estimated
on the fly using conditional source term
estimation~\cite{JinGB08}. Once the conditional rate is obtained,
through whatever means, the unconditional mean that appears in the
source term for the transport equations can be recovered by weighting
with the distribution and integrating over the conditioning space:
\begin{align}\label{eq:convolution}
  \wt{\dot{\omega}} = \int \langle \dot{\omega} | Z, c \rangle P(Z,c | \wt{Z}, \wt{Z''}, \wt{c}, \wt{c''}) \ud Z \ud c.
\end{align}
Here, $\langle \cdot \rangle$ denotes the volumetric mean of a
quantity; $\wt{\cdot} = \nicefrac{\ol{\rho~\cdot}}{\ol{\rho}}$ denotes
the Favre filter; $\ol{\cdot}$ denotes the \gls{les} filter; $Z$ is
the mixture fraction, capturing the mixing of fuel and oxidizer; $c$
is the progress variable, capturing the overall reaction progress;
$\dot{\omega}$ is the reaction rate of the progress variable (units of
$\unitfrac{1}{s}$, omitted for brevity); $Z'' = (Z-\wt{Z})^2$ is the
square of the mixture fraction subgrid scale fluctuation;
$c'' = (c - \wt{c})^2$ is the square of the progress variable subgrid
scale fluctuation; and $P(Z,c | \wt{Z}, \wt{Z''}, \wt{c}, \wt{c''})$
is the density-weighted \gls{pdf} of $Z$ and $c$, conditioned on
$\wt{Z}$, $\wt{Z''}$, $\wt{c}$, and $\wt{c''}$. It should be noted
that the \gls{ml} models presented in this study are being trained
using realizations of the \gls{fpdf}, referred to as \glspl{fdf},
computed from the DNS data and which are characterized by the subgrid
means and variances. This distinction between \glspl{fpdf} and
\glspl{fdf} will be adhered to throughout this work and follows the
convention presented by~\citet{Fox2003, Pitsch2006a}. The objective of
this work is to develop accurate models to generate \glspl{fpdf} for
\gls{les} using \gls{ml} techniques trained on \glspl{fdf} from the
\gls{dns} data set. Current analytical models often rely on using a
$\beta$ \gls{pdf}~\cite{Cook1994}. Though $\beta$-$\beta$ model is
based on a physically satisfying limiting behavior and is an
established presumed PDF model, the model is not universal and there
is ongoing research to improve the presumed PDF modeling
approach~\cite{Grout2009,Isaac2014,Linse2014}. The $\beta$ \gls{pdf} is defined as:
\begin{align}
  \label{eq:beta}
  \beta(x; a, b) = \frac{\Gamma(a + b)}{\Gamma(a)\Gamma(b)} x^{a-1} (1-x)^{b-1},
\end{align}
where $\Gamma(\cdot)$ is the gamma function; $a$ and $b$ are the
$\beta$ \gls{pdf} parameters, which can be related to the mean, $\mu$,
and variance, $\sigma^2$, as
$a=\mu \left( \frac{\mu (1-\mu)}{\sigma^2} - 1\right)$, and
$b=(1-\mu) \left( \frac{\mu (1-\mu)}{\sigma^2} - 1\right)$. In this
work, $\wt{Z}$ and $\wt{Z''}$ are used as the mean and variance for a
$\beta$ \gls{pdf} in the mixture fraction space, and $\wt{c}$ and
$\wt{c''}$ form the $\beta$ \gls{pdf} in the progress variable space,
such that
$P(Z,c) = \beta(Z; a_{\wt{Z}}, b_{\wt{Z}}) \beta(c; a_{\wt{c}},
b_{\wt{c}})$. The form of this expression for the $\beta$-$\beta$
model was chosen to be the product of two marginal $\beta$ PDFs
because it is the simplest closed-form analytical expression resulting
from using the first and second moments of the input variables. Other
expressions, such as the ``statistically most likely
distribution''~\cite{Ihme2008} or the generalized Dirichlet
distribution approach, could have been chosen though these result in
unclosed analytical forms requiring the solution of non-linear
equations at each grid point. Furthermore, following the
Connor-Mosimann approach and assuming unit-square support for the two
input variables results in the same product of two marginal $\beta$
PDFs chosen for this work~\cite{Perry2018}. Therefore, this model will
be used for comparisons with data-driven models using different
\gls{ml} techniques.

\subsection{Description of the \acrlong{dns} of the low-swirl burner}\label{sec:dns}
\begin{figure}[!tbp]%
  \centering%
  \includegraphics[width=0.5\textwidth]{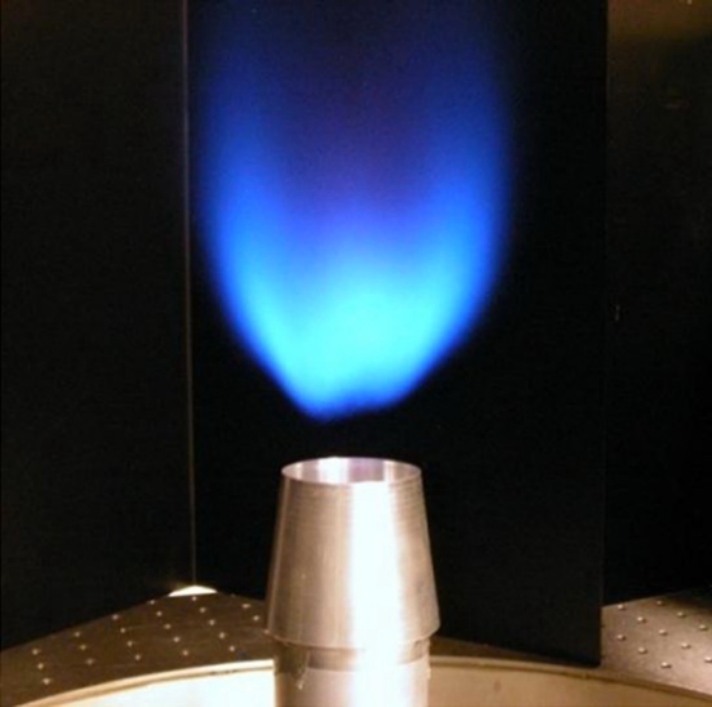}\\%
  \caption{Lifted flame of the low-swirl burner from the experimental configuration. From \citet{Day2012}. \copyright{} Elsevier. Reproduced with permission.}\label{fig:exp_lsb}%
\end{figure}

The \gls{dns} of an experimental lean premixed turbulent low-swirl
methane flame provide the data for model
development~\cite{Day2012,Cheng2000}. In this configuration, a nozzle
imposes a low swirl (geometric swirl number of 0.55) to a CH$_4$ and
air mixture with a fuel-air equivalence ratio of 0.7 at the inflow, Figure\,\ref{fig:exp_lsb}. A
co-flow of cold air surrounds the nozzle region with an upward
velocity of $0.25\,\unitfrac{m}{s}$. The inflow velocity of the
fuel-air mixture at the nozzle is $15\,\unitfrac{m}{s}$. The laminar
flame thickness is $600\,\unit{\mu m}$. The simulation was performed
using LMC, a low Mach number Navier-Stokes solver for turbulent
reacting flows that leverages adaptive mesh refinement to resolve
finer scales~\cite{Day2000}. Three levels of refinement were used,
leading to effective resolution of $100\,\unit{\mu m}$ in the flame
region. The computational domain was $0.25\,\unit{m}$ in each
dimension. The DRM 19 chemical mechanism was used to model the finite
rate kinetics~\cite{Kazakov1994}. The domain pressure is
$1\,\unit{atm}$. The physical characteristics and the mechanisms behind the flame are discussed in detail by~\citet{Day2012} and will not be described here for brevity. 

In the current analysis, the mixture fraction, $Z$, is computed
through a linear combination of the nitrogen mass fraction in the
burner exit stream and the co-flow and it is normalized such that it
varies between 0 in the co-flow stream and 1 in the burner exit
stream. The mixture fraction variable is used to quantify the mixing
between the stream from the fuel nozzle and the co-flow. When defined
this way, the mixture fraction variable has no relation to the local
fuel mass fraction and is a passive scalar with a transport equation
comprising only of temporal, advection, diffusion, and sub-grid
turbulent mixing terms. The progress variable is defined in this work
as $c= Y_{CO_2} + Y_{CO} + Y_{H_2} + Y_{H_2O}$ and varies between 0
and 0.21, where $Y_i$ is the mass fraction of species $i$,
$\sum^{N_s}_{i=1} Y_i = 1$, and $N_s$ is the number of species. This
definition of progress variable leads to a simpler transport equation
for the progress variable than a temperature based progress variable
or any other species mass fractions based progress variable. This work
is focused on determining models for the joint \gls{fdf} and, thus,
does not require the independence of the mixture fraction and the
progress variable.

\begin{figure}[!tbp]%
  \centering%
  \begin{subfigure}[t]{0.48\textwidth}%
    \includegraphics[page=1,width=\textwidth, trim = 1.2cm 0cm 0.4cm 0cm, clip=true]{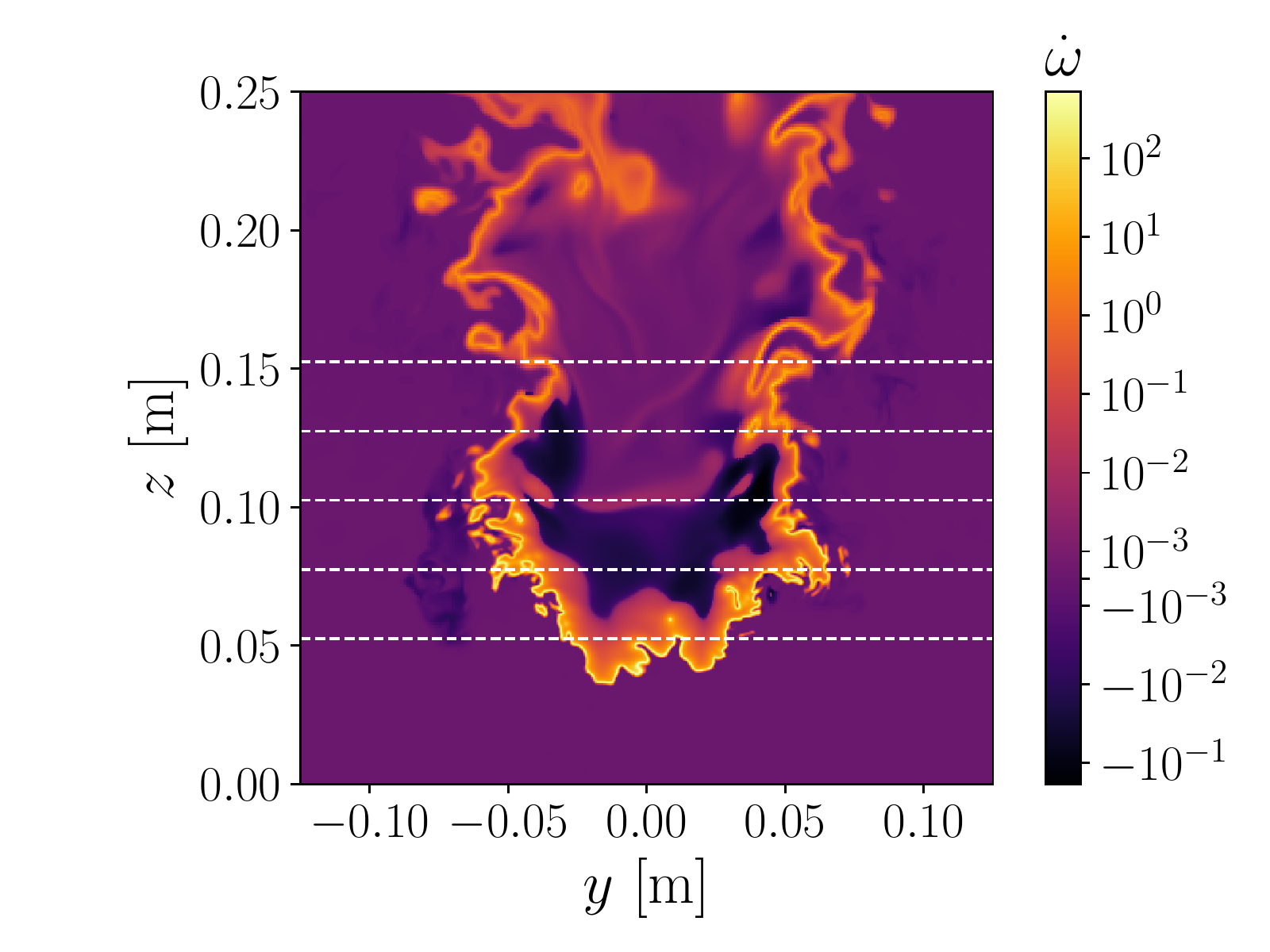}%
    \caption{$x=0\,\unit{m}$.}%
  \end{subfigure}\hfill%
  \begin{subfigure}[t]{0.48\textwidth}%
    \includegraphics[page=2,width=\textwidth, trim = 1.2cm 0cm 0.4cm 0cm, clip=true]{./figs/src_pv.pdf}%
    \caption{$z=0.0525\,\unit{m}$ (center height of $\mathcal{V}_{1}$).}%
  \end{subfigure}\\%
  \begin{subfigure}[t]{0.48\textwidth}%
    \includegraphics[page=3,width=\textwidth, trim = 1.2cm 0cm 0.4cm 0cm, clip=true]{./figs/src_pv.pdf}%
    \caption{$z=0.0775\,\unit{m}$ (center height of $\mathcal{V}_{3}$).}%
 \end{subfigure}\hfill%
  \begin{subfigure}[t]{0.48\textwidth}%
    \includegraphics[page=4,width=\textwidth, trim = 1.2cm 0cm 0.4cm 0cm, clip=true]{./figs/src_pv.pdf}%
    \caption{$z=0.1025\,\unit{m}$ (center height of $\mathcal{V}_{5}$).}%
  \end{subfigure}\\%
  \begin{subfigure}[t]{0.48\textwidth}%
    \includegraphics[page=5,width=\textwidth, trim = 1.2cm 0cm 0.4cm 0cm, clip=true]{./figs/src_pv.pdf}%
    \caption{$z=0.1275\,\unit{m}$ (center height of $\mathcal{V}_{7}$).}%
  \end{subfigure}\hfill%
  \begin{subfigure}[t]{0.48\textwidth}%
    \includegraphics[page=6,width=\textwidth, trim = 1.2cm 0cm 0.4cm 0cm, clip=true]{./figs/src_pv.pdf}%
    \caption{$z=0.14\,\unit{m}$ (center height of $\mathcal{V}_{9}$).}%
  \end{subfigure}
  \caption{Slices of $\dot{\omega}$ in \gls{dns}. White dashed lines: $z$ locations of slices shown in (b)--(f).}\label{fig:dns}%
\end{figure}

The primary motivation behind choosing this test case to demonstrate
the capabilities of \gls{ml} for joint \gls{fdf} models is the
presence of multiple burning regimes. As seen in
Figure\,\ref{fig:dns}, a premixed flame lifted from the fuel nozzle is
clearly observed. The products from this premixed flame mix downstream
with the air from the co-flow, which sets up a secondary reacting
zone. A number of modeling challenges are introduced because of the
presence of these multiple regimes. One of these modeling challenges
is to capture the joint \gls{fdf} describing the mixing of the mixture
fraction, a passive scalar, with the progress variable, an active
scalar. Analytical models for joint \glspl{fdf} have found to be
lacking accuracy for such complex configurations~\cite{Ihme2008}, and,
this motivates the exploration of \gls{ml} techniques for constructing
models for challenging turbulent combustion problems, such as the
configuration considered in this study. Additional advantages of using
data from multiple burning regimes for training \gls{ml} models include
using of more diverse training data, avoiding overfitting, and
increasing the opportunities for model generalization.

\begin{figure}[!tbp]%
  \centering%
  \begin{subfigure}[t]{0.48\textwidth}%
    \includegraphics[page=1,width=\textwidth, trim=0.4cm 0cm 1.4cm 0cm, clip=true]{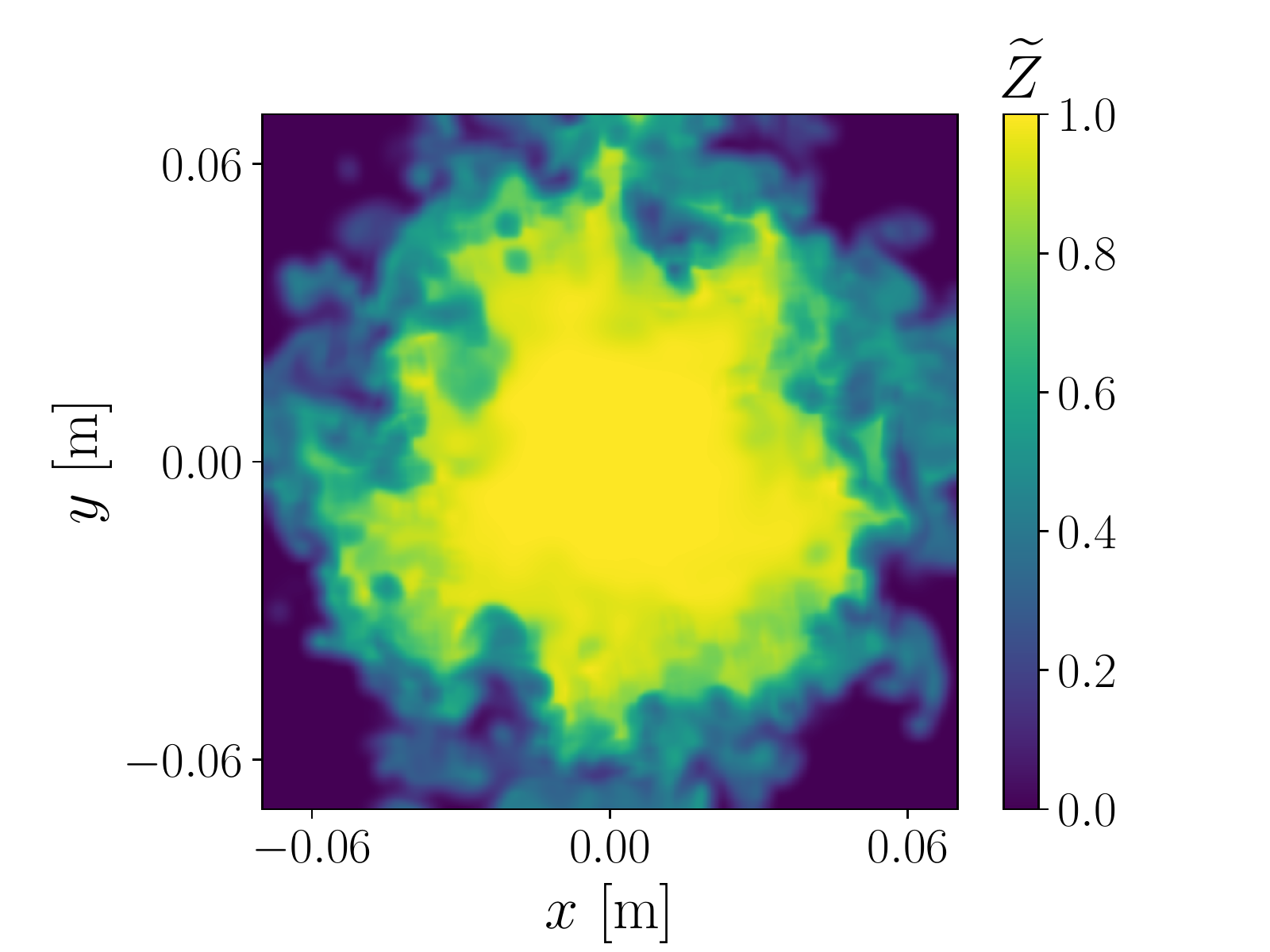}%
    \caption{$\wt{Z}$.}%
  \end{subfigure}\hfill%
  \begin{subfigure}[t]{0.48\textwidth}%
    \includegraphics[page=2,width=\textwidth, trim=0.4cm 0cm 1.4cm 0cm, clip=true]{./figs/dice_0004_slice.pdf}%
    \caption{$\wt{Z''}$.}%
  \end{subfigure}\\%
  \begin{subfigure}[t]{0.48\textwidth}%
    \includegraphics[page=3,width=\textwidth, trim=0.4cm 0cm 1.4cm 0cm, clip=true]{./figs/dice_0004_slice.pdf}%
    \caption{$\wt{c}$.}%
 \end{subfigure}\hfill%
  \begin{subfigure}[t]{0.48\textwidth}%
    \includegraphics[page=4,width=\textwidth, trim=0.4cm 0cm 1.4cm 0cm, clip=true]{./figs/dice_0004_slice.pdf}%
    \caption{$\wt{c''}$.}%
  \end{subfigure}
  \caption{Slices of filtered \gls{dns} data at $z=0.0775\,\unit{m}$ (center height of $\mathcal{V}_3$).}\label{fig:slices}%
\end{figure}%

\subsection{Generation of the modeling data}\label{sec:gen}

A data set of sample moments and associated \glspl{fdf} was generated
from a statistically stationary single time snapshot at
$t=0.0626\,\unit{s}$ from this \gls{dns} by considering different
sub-volumes of the domain that span the flame, from the region of
premixed burning of the fuel-air mixture from the nozzle to the mixing
zone between the products from the primary premixed flame and the air
from the co-flow. These volumes --- denoted by $\mathcal{V}_i$, where
$i=1, \dots, n_v$ and $n_v = 9$ is the number of sub-volumes --- are
centered at $z_i = 0.0525\,\unit{m} + (i-1) 0.0125 \,\unit{m}$, with
height $0.00625\,\unit{m}$ and width $0.14\,\unit{m}$, composed of
$1146 \times 1146 \times 51$ cells. The locations of several of these
subregions and planar slices of $\dot{\omega}$ are presented in
Figure\,\ref{fig:dns}. The premixed lifted flame, with high values of
$\dot{\omega}$ and steep gradients corresponding to a thin flame, can
be observed around $z=0.05\,\unit{m}$. Farther downstream of the
nozzle, the premixed flame products mix with the air coming from the
co-flow and react to produce lower values of
$\dot{\omega}$. Representative slices of the filtered \gls{dns} data
in $\mathcal{V}_3$ are presented in Figure\,\ref{fig:slices}. The core
of the flame is fully burned as seen by high values of $\wt{c}$, and
the reactions take place in a thin region at the interface of the
fuel-air mixture from the nozzle and co-flow air.

Throughout this work, samples refer to a pointwise sampling of the
filtered fields, each with an associated collection of moments and an
\gls{fdf}; volumes refer to a subset of the samples divided according
to regions of the domain; and the \gls{fdf} for each sample is
described by the four sample moments,
$\left[\wt{Z}, \wt{Z''}, \wt{c}, \wt{c''}\right]$. In each volume,
sample moments and associated \glspl{fdf} were generated by
using a discrete box filter:
\begin{align}
  \label{eq:box}
  \ol{\phi} (x, y, z) = \frac{1}{n_f^3}\sum_{i=-\nicefrac{n_f}{2}}^{\nicefrac{n_f}{2}} \sum_{~j=-\nicefrac{n_f}{2}}^{\nicefrac{n_f}{2}} \sum_{~k=-\nicefrac{n_f}{2}}^{\nicefrac{n_f}{2}} \phi(x + i \Delta x, y + j \Delta x, z + k \Delta x) 
\end{align}
where $\phi$ is the variable to be filtered,
$n_f = \nicefrac{\ol{\Delta}}{\Delta}$ is the number of points in the
discrete box filter, $\ol{\Delta} = 32 \Delta x$ is the filter length
scale, and $\Delta x = 100 \,\unit{\mu m}$ is the smallest spatial
discretization in the \gls{dns} (six times smaller than the laminar
flame thickness). The filter length scale was chosen to be
representative of typical \gls{les} filter scales~\cite{Pitsch2006a}
and to ensure an adequate sampling of the \gls{fdf} at the filter
scale. These filters were equidistantly spaced at $8\Delta x$, leading
to 58800 \glspl{fdf} for each volume. The computed
conditional \glspl{fdf} are the density-weighted
\glspl{fdf} of $Z$ and $c$, discretized with 64 bins in $Z$ and 32
bins in $c$. For notational convenience, $P(Z,c) = P(Z = Z^*, c=c^*)$
will be used in this work, and the discrete density functions will be
referred to as density functions instead of mass functions. The
conditional means of the reaction rate,
$\langle \dot{\omega} | Z, c \rangle$, are also computed for each
sample with an identical discretization.

\begin{figure}[!tbp]%
  \centering%
  \begin{subfigure}[t]{0.48\textwidth}%
    \includegraphics[page=9,width=\textwidth, trim=0.5cm 0cm 1.5cm 1.1cm, clip=true]{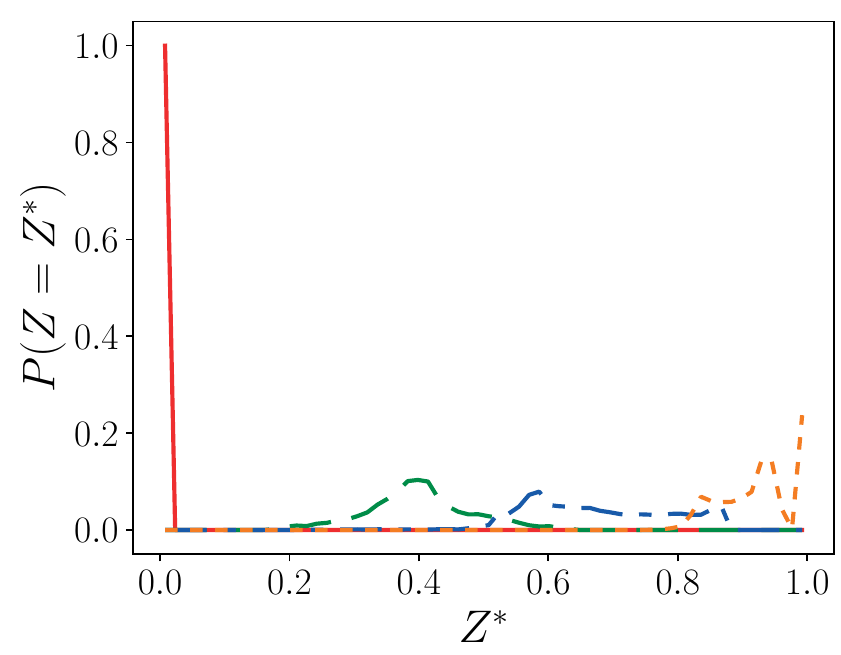}%
    \caption{Marginal $P(Z,c)$ as a function of $Z$.}%
  \end{subfigure}\hfill%
  \begin{subfigure}[t]{0.48\textwidth}%
    \includegraphics[page=10,width=\textwidth, trim=0.5cm 0cm 1.5cm 1.1cm, clip=true]{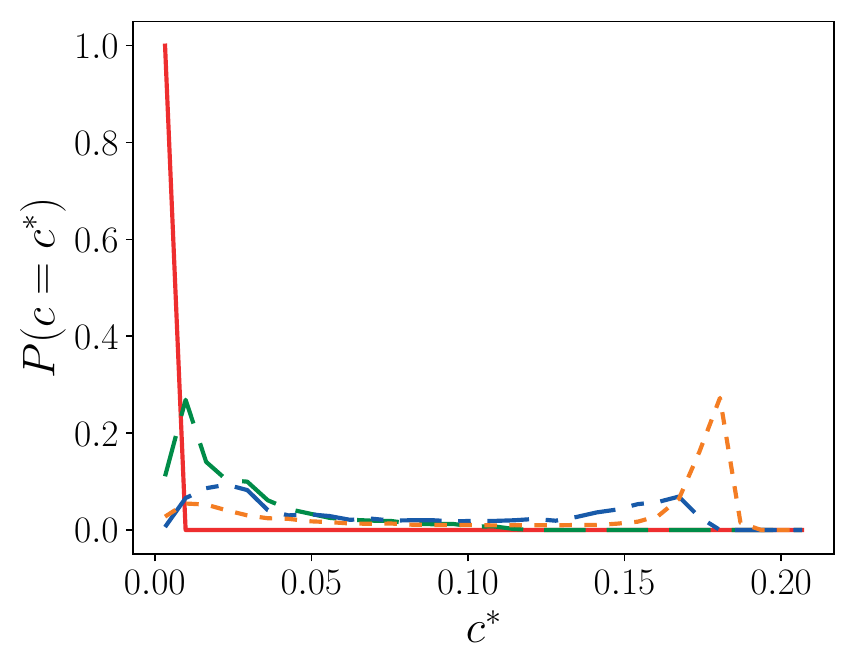}%
    \caption{Marginal $P(Z,c)$ as a function of $c$.}%
  \end{subfigure}\\%
  \begin{subfigure}[t]{0.48\textwidth}%
    \includegraphics[page=11,width=\textwidth, trim=0.5cm 0cm 1.5cm 1.1cm, clip=true]{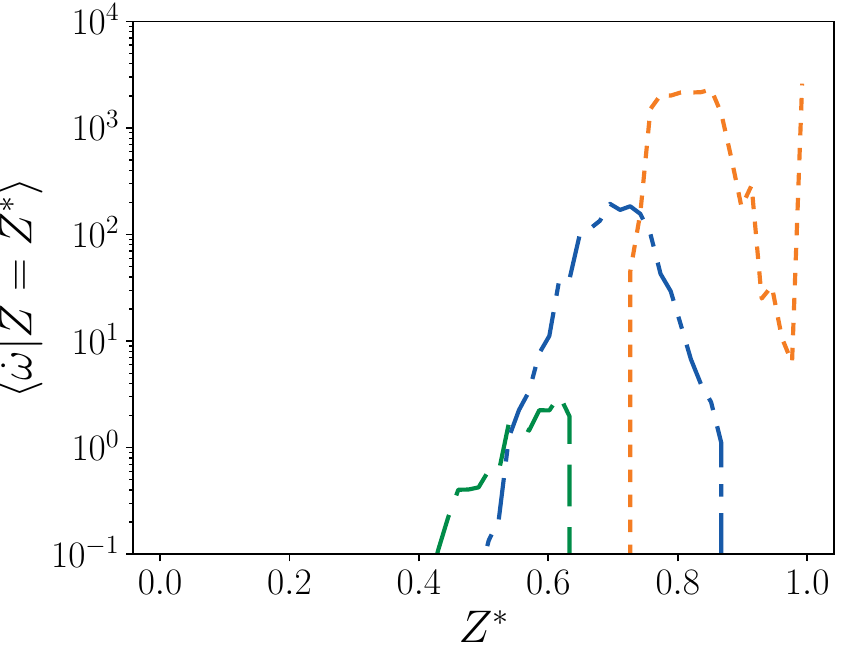}%
    \caption{Marginal conditional means of $\dot{\omega}$ as a function of $Z$.}%
  \end{subfigure}\hfill%
  \begin{subfigure}[t]{0.48\textwidth}%
    \includegraphics[page=12,width=\textwidth, trim=0.5cm 0cm 1.5cm 1.1cm, clip=true]{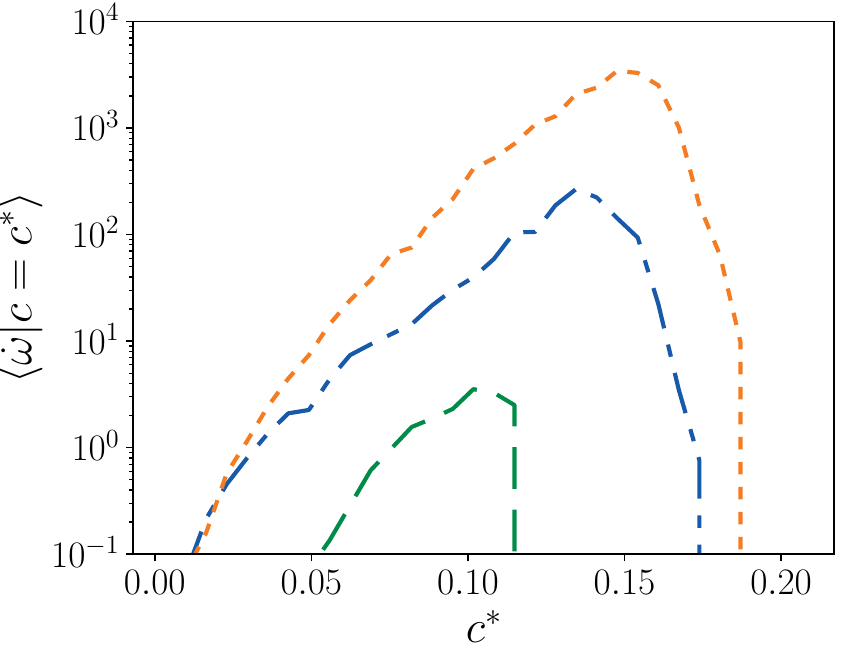}%
    \caption{Marginal conditional means of $\dot{\omega}$ as a function of $c$.}%
  \end{subfigure}%
  \caption{Examples of $P(Z,c)$ and $\langle \dot{\omega} | Z, c \rangle$ for increasing $\wt{\dot{\omega}}$. Red solid: $\wt{\dot{\omega}} = 0$ ($\wt{Z} = 0$, $\wt{Z''} = 0$, $\wt{c} = 0$, $\wt{c''} = 0$); green dashed: $\wt{\dot{\omega}} = 0.03$ ($\wt{Z}  =0.4$, $\wt{Z''}=0.006$, $\wt{c}  =0.03$, $\wt{c''}=0.0006$); blue dash-dotted: $\wt{\dot{\omega}} = 7.4$ ($\wt{Z}  =0.7$, $\wt{Z''}=0.01$, $\wt{c}  =0.08$, $\wt{c''}=0.003$); orange short dashed: $\wt{\dot{\omega}} = 42.2$ ($\wt{Z}  =0.9$, $\wt{Z''}=0.003$, $\wt{c}  =0.12$, $\wt{c''}=0.005$).}\label{fig:pdfs}%
\end{figure}

Examples of $P(Z,c)$ and $\langle \dot{\omega} | Z, c \rangle$ in
$\mathcal{V}_3$ for increasing $\wt{\dot{\omega}}$ illustrate the wide
range of observed shapes, Figure\,\ref{fig:pdfs}. For high
$\wt{\dot{\omega}}$, the conditional means of $\dot{\omega}$ peak at
$c=0.16$ and exhibit a bimodal distribution at high $Z$ because of the
burning of the fuel stream from the nozzle ($Z=1$) and the burning of
the products mixing with the co-flow. For intermediate
$\wt{\dot{\omega}}$, the conditional means of $\dot{\omega}$ are
largest at $Z=0.7$ and $c=0.14$, which is also attributed to the
burning of the mixed products. As $\wt{\dot{\omega}}$ increases, the
location of the peak of $P(Z, c)$ increases in the $Z$ and $c$ space
because reactions happen at higher $Z$ and $c$.

\begin{figure}[!tbp]%
  \centering%
  \includegraphics[page=1, height=0.49\textwidth, trim=0.0cm 0cm 6.cm 0cm, clip]{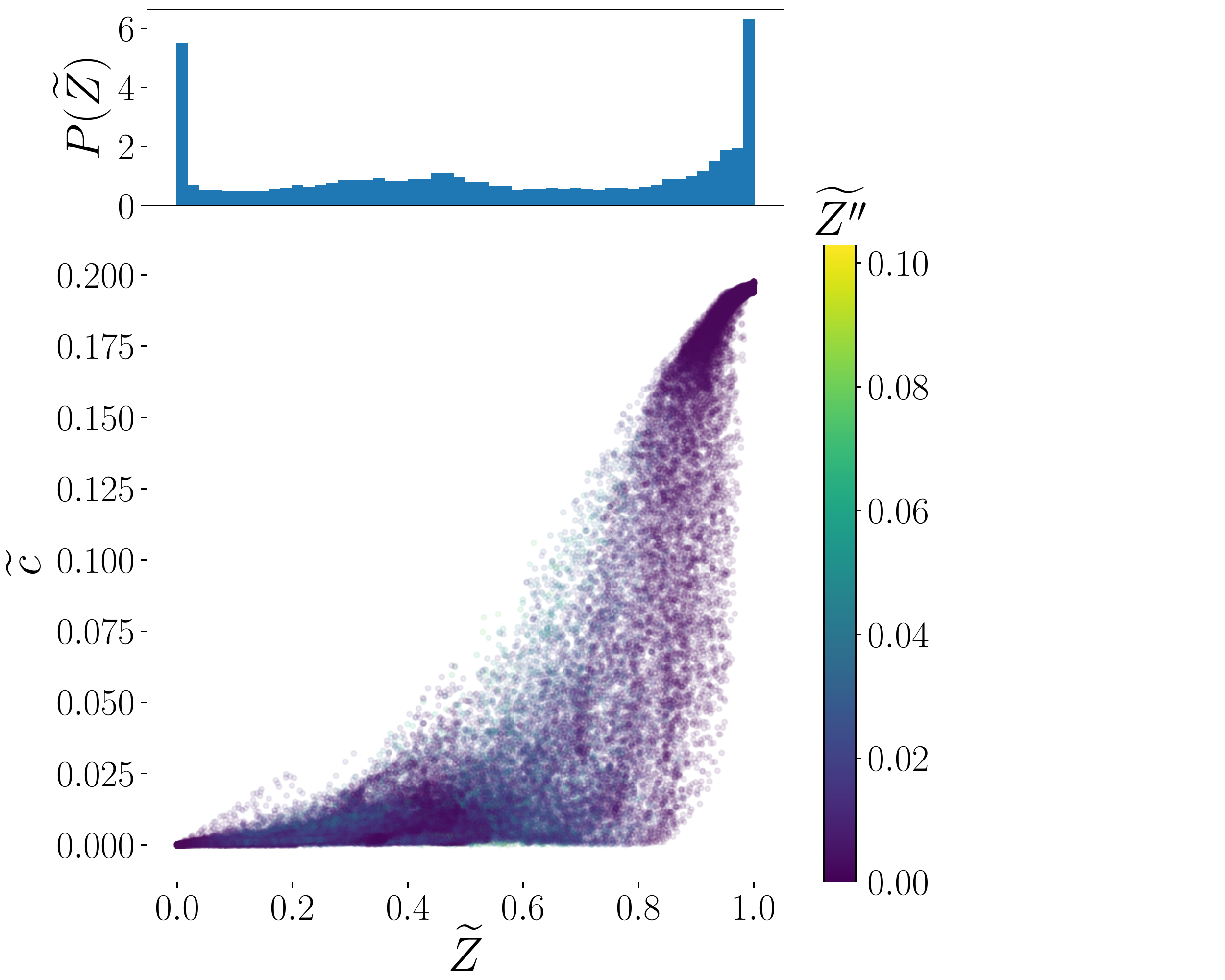}\hfill%
  \includegraphics[page=2, height=0.49\textwidth, trim=1.8cm 0cm 1.8cm 0cm, clip]{./figs/inputs_dice_0004.pdf}%
  \caption{Scatter plots of moments $\wt{Z}$ and $\wt{c}$ for samples in $\mathcal{V}_3$ (centered at $z=0.0775\,\unit{m}$) colored by $\wt{Z''}$ (left) and $\wt{c''}$ (right) with associated marginal distributions.}\label{fig:inputs}%
\end{figure}

Figure\,\ref{fig:inputs} illustrates the distribution of moments,
$\left[\wt{Z}, \wt{Z''}, \wt{c}, \wt{c''}\right]$, across the samples
in $\mathcal{V}_3$. Most \glspl{fdf} are associated with fully burned
states originating from the premixed burning of the fuel-air mixture
from the nozzle ($\delta$ \glspl{pdf} centered at $\left( \wt{Z}, \wt{c} \right) = (1,0.2)$) or the nonreacting unburned states (
$\delta$ \glspl{pdf} centered at
 and $(0,0)$). A significant number of the \glspl{fdf},
however, are associated with intermediate states spanning the full
range of $\wt{Z}$ and $\wt{c}$ with larger $\wt{Z''}$ and $\wt{c''}$
because of the burning of the products from the primary premixed flame
zone mixed with the air from the co-flow. 

\subsection{Machine learning algorithms}\label{sec:methods}
In this work, we evaluate the performance and suitability of three
different types of \gls{ml} algorithms, each representative of a
prevalent class in \gls{ml}: (i) random forest for traditional
\gls{ml}, (ii) feed-forward \gls{dnn} for deep learning, and (iii)
\gls{cvae} for generative and unsupervised learning. The model
hyperparameters are summarized in~\ref{app:hp_summary}.

\begin{figure}[!tbp]%
  \centering%
  \includegraphics[width=0.9\textwidth]{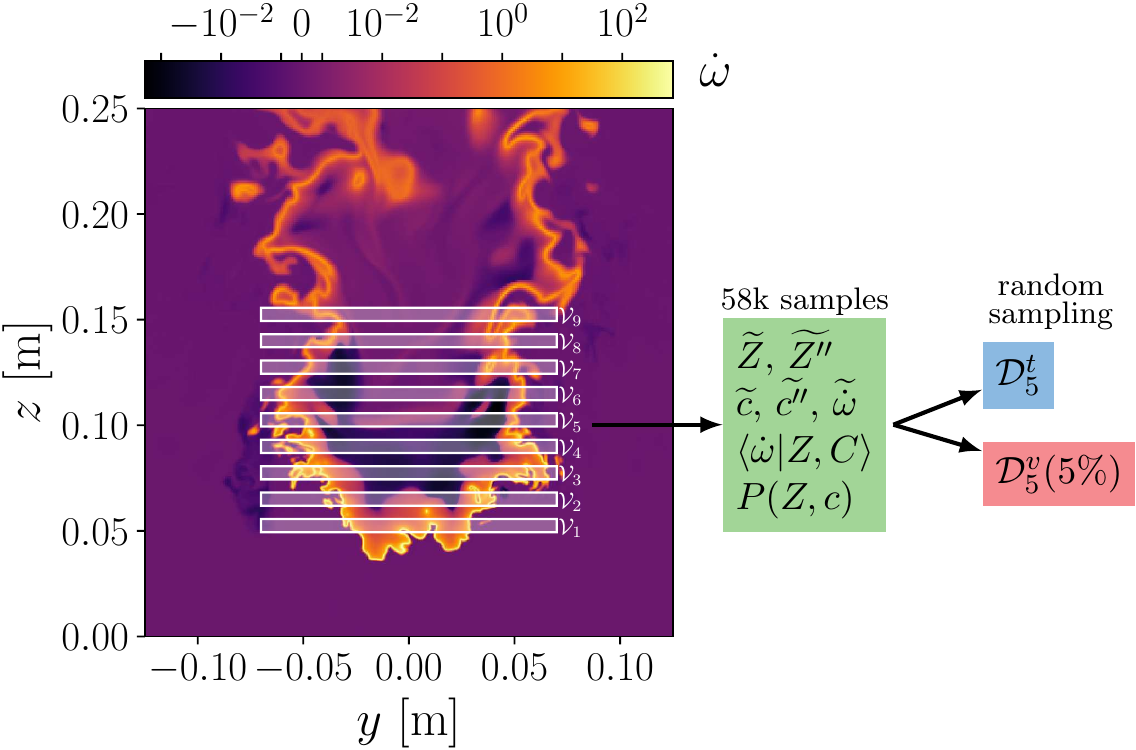}\\%
  \caption{Illustration of data generation procedure for $\mathcal{V}_5$.}\label{fig:gen_data}%
\end{figure}

The model inputs are the four sample moments,
$\left[ \wt{Z}, \wt{Z''}, \wt{c}, \wt{c''} \right]$, and the outputs
are the 2048 discrete points representing the joint \gls{fdf} ($64$ in
$Z$, $32$ in $c$). The samples from a volume,
$\mathcal{V}_i~(i=1,\dots, n_v)$, are randomly distributed among two
distinct data sets: a training data set, $\mathcal{D}_i^t$, used to
train the algorithms; and a validation data set, $\mathcal{D}_i^v$,
used to validate the algorithms and comprising $5\%$ of the samples,
i.e.,\ $|\mathcal{D}_i^v| = 2940$, where $|\cdot|$ denotes the
cardinality of the data set. Figure\,\ref{fig:gen_data} illustrates
this process for $\mathcal{V}_5$. In this work, we evaluate different
models using different training strategies:
\begin{enumerate}
\item Models trained using $\mathcal{D}_3^t$ and evaluated on $\mathcal{D}_i^v$ ($i=1, \dots, n_v$);
\item Models trained using $\mathcal{D}_5^t$ and evaluated on $\mathcal{D}_i^v$ ($i=1, \dots, n_v$);
\item Models trained using $\mathcal{D}^t = \bigcup\limits_{i=1, 3, 5, 7, 9} \mathcal{D}_i^t$ and evaluated on $\mathcal{D}_i^v$ ($i=1, \dots, n_v$).
\end{enumerate}
The first two strategies involve training and validating on different
physical regions of the flame. The third strategy uses training data
from the entire flame and the validation data from the training
regions and intermediate regions. Prior to training, the sample
moments were independently scaled by subtracting the median and
dividing the data by the range between the $25^\text{th}$ and
$75^\text{th}$ quantiles. This scaling is robust to
outliers~\cite{Pedregosa2011}. A separate scaling was computed for
each training data set and applied to the associated validation data
set. The evaluation of a model $m$ on a data set $\mathcal{D}$ is
denoted $m(\mathcal{D})$.

\begin{figure}[!tbp]%
  \centering%
  \begin{subfigure}[t]{0.48\textwidth}%
    \centering
    \includegraphics[width=0.85\textwidth]{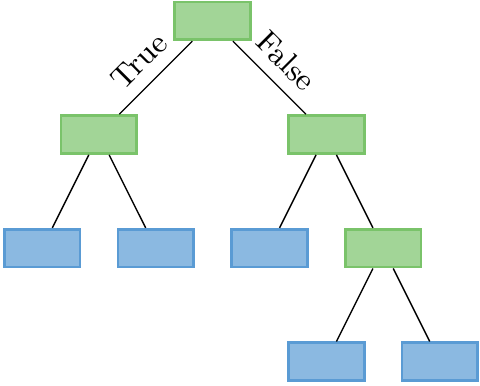}%
    \caption{Decision tree with a tree depth of 3. Green boxes: decision nodes with conditions on the input variables; blue boxes: leaves with output values.}\label{fig:rf}%
  \end{subfigure}\hfill%
  \begin{subfigure}[t]{0.48\textwidth}%
    \centering
    \includegraphics[width=\textwidth]{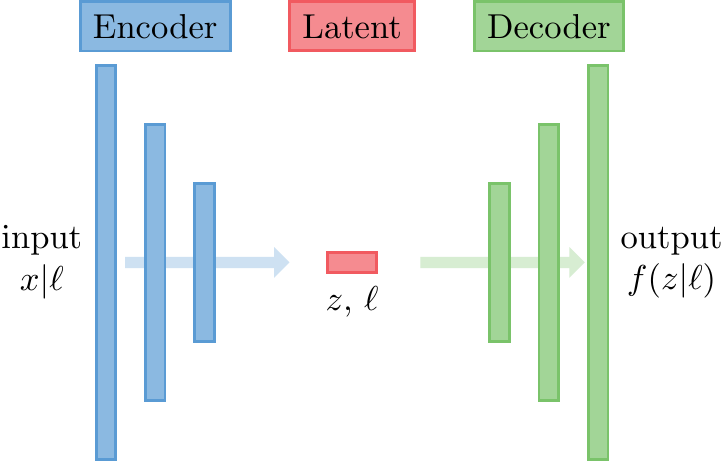}%
    \caption{\Gls{cvae} architecture with input $x$, labels $\ell$, and latent space $z$.}\label{fig:cvae}%
  \end{subfigure}%
  \caption{Diagrams of \gls{ml} algorithm architectures.}
\end{figure}

The first of the investigated models, random forests (RF), is an
ensemble model that creates ensembles of low-bias/high-variance
individual decision trees and uses the average of the individual model
predictions to provide the prediction for the overall
forest~\cite{Breiman2001}. A decision tree is a model that uses a
treelike structure to represent nodes that encode conditions based
on the input variables, branches that split from each node, and
termination points, i.e.,\ leaves, which provide the target value
predictions, Figure\,\ref{fig:rf}. The main parameter for a decision
tree model is the maximum tree depth, which is the length of the
longest path from the root of the tree to a leaf.

Two key insights have driven the effectiveness of random forests
models for complex tasks while avoiding
overfitting~\cite{Fernandez-Delgado2014,Liaw2002}, a problem arising
when a model is overly accurate on the training data while failing to
predict non-training data. The first is that it leverages bootstrap
aggregating or bagging, a method that improves model stability,
accuracy, and overfitting problems by dividing the training set into
several smaller training sets, called bootstraps, populated through
random uniform sampling with replacement. In random forests, each
decision tree is built using a different bootstrap of the training
data. The second is that, instead of splitting each node in the tree
according to the best split of all the variables, the split is done
using the best split among a random subset of the variables. The two
key parameters of the random forests algorithm are the number of
decision trees and the depth of the decision trees. For this work, the
random forests model contains 100 decision trees and a maximum tree
depth of 30 nodes, beyond which results were insensitive to the model
size, and the model size grew larger than can be effectively trained
on what we consider a typical analysis workstation with 256 GB of
memory. The total model \gls{dof}, measured as the sum of nodes in
each tree, is 5.2 million. Though no constraints were explicitly
imposed on the outputs of the random forest model, the model
predictions exhibited properties of \glspl{pdf} (integration to unity
and bounded between zero and one).

The field of deep learning has exhibited success in developing models
for tasks ranging from image
processing~\cite{Goodfellow2014,Burger2012,Dosovitskiy2015,Lefkimmiatis2016,Ledig2017,Tai2017,Lai2017}
to text generation~\cite{Graves2013,Wu2016,Kwon2017} and
games~\cite{Silver2017}. Several reviews of the field give a summary
of recent breakthroughs and developments~\cite{Lecun2015,
  Schmidhuber2015, Prieto2016, Goodfellow2016, Liu2017}. As a first
example of deep learning, we develop a feed-forward, fully connected
\gls{dnn} for presumed \gls{pdf} modeling. Similar to the decoder
network presented below, this network consists of two hidden layers
and an output layer. The hidden layers comprise,
respectively, 256, and 512 fully connected nodes, a leaky rectified
linear unit activation function:
\begin{align}
  \label{eq:relu}
  y = R(x) =
  \begin{cases}
    x, & \text{ if } x \geq 0, \\
    \alpha x, & \text{ otherwise, }
  \end{cases}
\end{align}
where $x$ is the layer input vector, $y$ is the layer output vector,
and $\alpha=10^{-2}$ is a small slope; and a batch normalization
layer~\cite{Ioffe2015}:
\begin{align}
  \label{eq:bn}
  y = B(x) = \gamma \frac{x - \mu_x}{\sqrt{\sigma_x^2 + \epsilon}} + \delta,
\end{align}
where $x$ is the layer input vector of size $n$, $y$ is the layer
output vector of size $n$, $\mu_x = \nicefrac{1}{n} \sum_{i=1}^n x_i$,
$\sigma_x^2 = \nicefrac{1}{n} \sum_{i=1}^n (x_i - \mu_x)^2$,
$\epsilon = 10^{-5}$, and $\gamma$ and $\delta$ are learnable
parameter vectors of the same size as $x$. For inference, i.e.,\
prediction on new data, the batch normalization layer uses a moving
average of $\mu_x$ and $\sigma_x$ with a decay of $0.1$ computed
during training. Because we are interested in predicting \glspl{pdf}, we apply a softmax activation function:
\begin{align}
  y = S(x) = \frac{\exp{(x)}}{\sum^n_{i=1} \exp{(x_i)}},
\end{align}
where $x$ is the layer input vector of size $n$, and $y$ is the layer
output vector of size $n$, on the output layer to ensure
that $\sum^n_{i=1} y_i = 1$ and
$y_i \in [0,1]~\forall i = 1, \dots, n$. Additionally, the loss
function for the network is the binary cross entropy between the
target, $t$, and the output, $y$:
\begin{align}
  l(y,t) = \frac{1}{n} \sum_{i=1}^n{\left( t_i \log{(y_i)}+(1-t_i) \log(1-y_i) \right)},
\end{align}
and is a good metric for measuring differences between
\glspl{pdf}. The total \gls{dnn} \gls{dof}, measured as the number of
trainable parameters, is 1.1 million. The training occurs during 500
epochs, where an epoch implies one training cycle through the entire
training data, after which the loss on the training data is
converged. For each epoch, the training data is fully shuffled and
divided into batches with 64 training samples per batch. The specific
gradient descent algorithm for this work is the Adam
optimizer~\cite{Kingma2014} with an initial learning rate of
$10^{-4}$. The learning rate is a dimensionless parameter that
determines the step size of the stochastic gradient descent used to adjust
the model weights of the neural network. The Adam optimizer presents
many more advantages than traditional stochastic gradient descent by
maintaining a per-parameter learning rate, which is adapted during
training based on exponential moving averages of the first and second
moments of the gradients. The network was implemented in
Pytorch~\cite{Paszke2017} and trained on a single NVIDIA Tesla K80
GPU.

Recently, deep generative algorithms, in the form of
\glspl{vae}~\cite{Kingma2013, Rezende2014} and
\glspl{gan}~\cite{Goodfellow2014}, have illustrated how encoding
features into a latent space can provide an accurate framework for
generating samples from a learned data distribution. Interpolation and
other operations in the latent space have shown success in generating
samples that usefully combine features of the data set. Because the
modeling challenge presented here presents physical regimes with
different combustion characteristics, this latent space representation
may be advantageous for interpolation and generalization. Though
supervision can be built into the network by adding labels to the
input and latent spaces, these algorithms are unsupervised learning
algorithms. The \gls{vae} relies on an encoder, decoder, and loss
function. The encoder transforms the input data into a latent
space. Unlike encoders for standard autoencoders, the encoder outputs
two vectors: a vector of means and a vector of standard
deviations. These form the parameters of the random normal variable to
be sampled in the latent space. This implies that, given the same
data, the encoding in the latent space will differ slightly on
different passes. The decoder transforms the resulting encoding in the
latent space into outputs that are designed, through the definition of
the loss function, to be generated samples from the same distribution
as the input data. The loss function is a negative log-likelihood
combined with a regularizer. The negative log-likelihood measures the
reconstruction loss by the decoder. The regularizer is the
Kullback-Leibler divergence between the encoder distribution and the
distribution in the latent space, thereby enforcing a continuous
latent space. The \gls{vae} used in this work follows an
hourglass-type architecture, Figure\,\ref{fig:cvae}. The encoder
network comprises an input layer with 2048 nodes, a hidden layer with
512 nodes, and the last hidden layer with 256 nodes. The decoder
network is a mirror image of the encoder (256, 512, and 2048 nodes in
each layer). The activation functions in the encoder and decoder are
rectified linear units. The final activation function in the decoder
is a softmax function, similar to the \gls{dnn}. The total \gls{dnn}
\gls{dof}, measured as the number of trainable parameters, is 2.3
million. The batch size for each epoch is 64, and the network was
trained for 500 epochs. The Adam optimizer was used with an initial
learning rate of $10^{-3}$. The latent space dimension is 10. We use a
minor variation of the \gls{vae} called the \gls{cvae}, allowing for
the conditioning of the input on a set of labels. The labels are
passed both to the encoder with the input data and to the decoder with
the latent space sample data. Therefore, unlike the two previous
models, the \gls{cvae} model input is the discrete exact \glspl{fdf}
and the four sample moments,
$\left[ \wt{Z}, \wt{Z''}, \wt{c}, \wt{c''} \right]$ (the sample
moments are also inputs for the latent space); and the \gls{cvae}
model output is the discrete modeled \gls{fdf}. For the \gls{fdf}
inference, the sample moments are combined with a latent space
sampling of a standard normal distribution and passed through the
decoder part of the \gls{cvae}. The network was implemented in
Pytorch~\cite{Paszke2017} and trained on a single NVIDIA Tesla K80
GPU.

Although a conditional \gls{gan} using the infoGAN network
architecture~\cite{Chen2016a} was evaluated for this work, it did not
perform as well as the \gls{cvae} because of difficulties related to the
stability of training a multi-agent model, and results from this model
are omitted for brevity.

\section{Results}\label{sec:results}

In this section, we present results of using the \gls{ml} techniques
to model the \gls{fpdf},
$P(Z,c | \wt{Z}, \wt{Z''}, \wt{c}, \wt{c''})$, from
Equation\,(\ref{eq:convolution}). We first focus on using data from
the volume centered at $z=0.1025\,\unit{m}$ because this section of
the domain contains regions that are dominated by premixed burning of
the fuel-air mixture from the nozzle and the burning of the products
from the primary premixed flame mixing with the air from the co-flow,
as discussed in Section\,\ref{sec:dns}. Next, we evaluate the
generalization capabilities of the different algorithms by
characterizing their performance on other sections of the flame.

\begin{figure}[!tbp]%
  \centering%
  \begin{subfigure}[t]{\textwidth}%
    \includegraphics[page=3,width=0.48\textwidth, trim=0.2cm 0cm 1.5cm 1.3cm, clip=true]{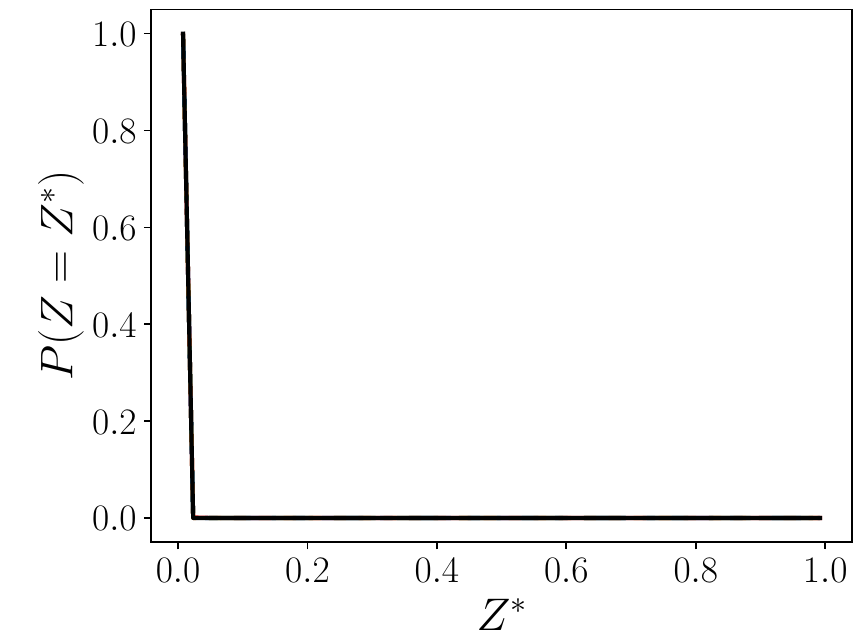}\hfill%
    \includegraphics[page=4,width=0.48\textwidth, trim=0.2cm 0cm 1.5cm 1.3cm, clip=true]{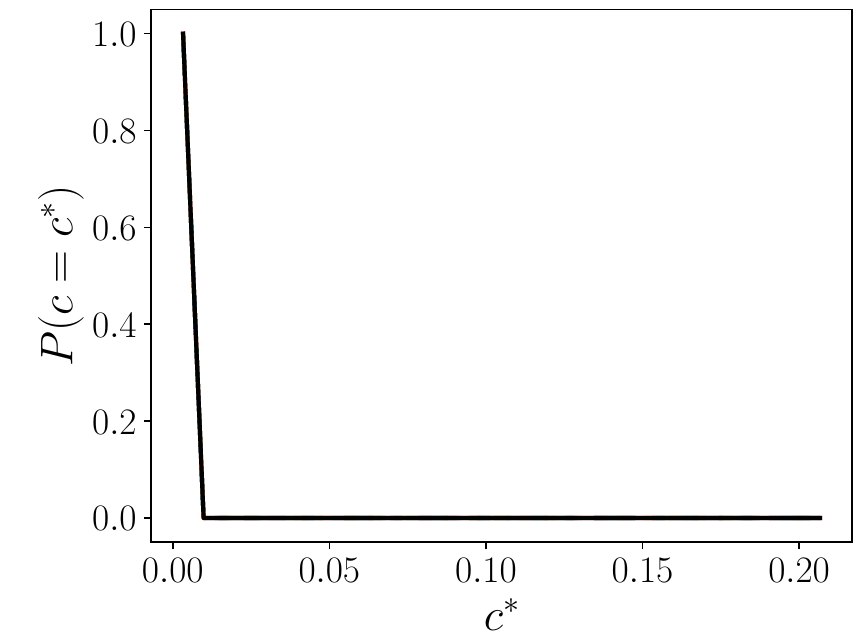}%
    \caption{$J(P||P_{\beta})=0$.}\label{fig:}%
  \end{subfigure}\hfill%
  \begin{subfigure}[t]{\textwidth}%
    \includegraphics[page=3,width=0.48\textwidth, trim=0.2cm 0cm 1.5cm 1.3cm, clip=true]{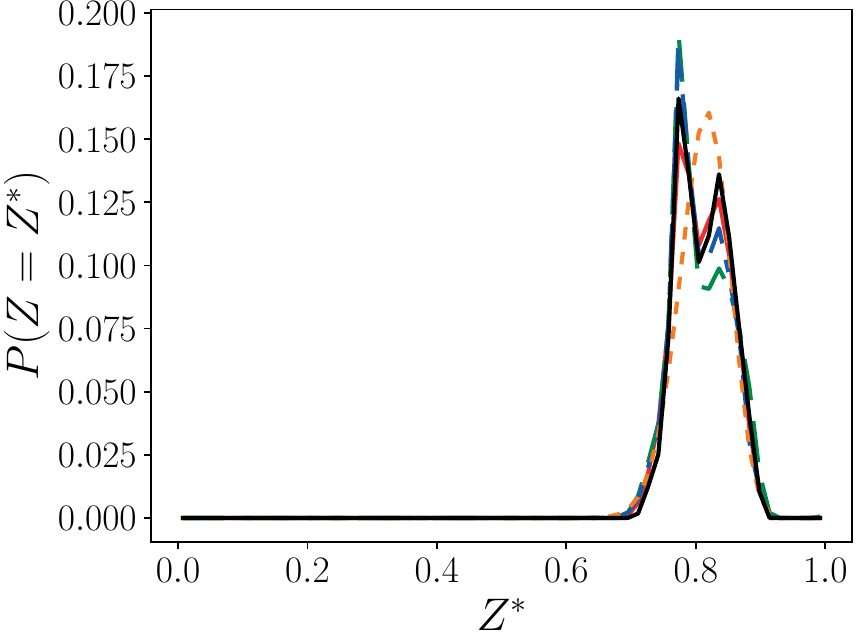}\hfill%
    \includegraphics[page=4,width=0.48\textwidth, trim=0.2cm 0cm 1.5cm 1.3cm, clip=true]{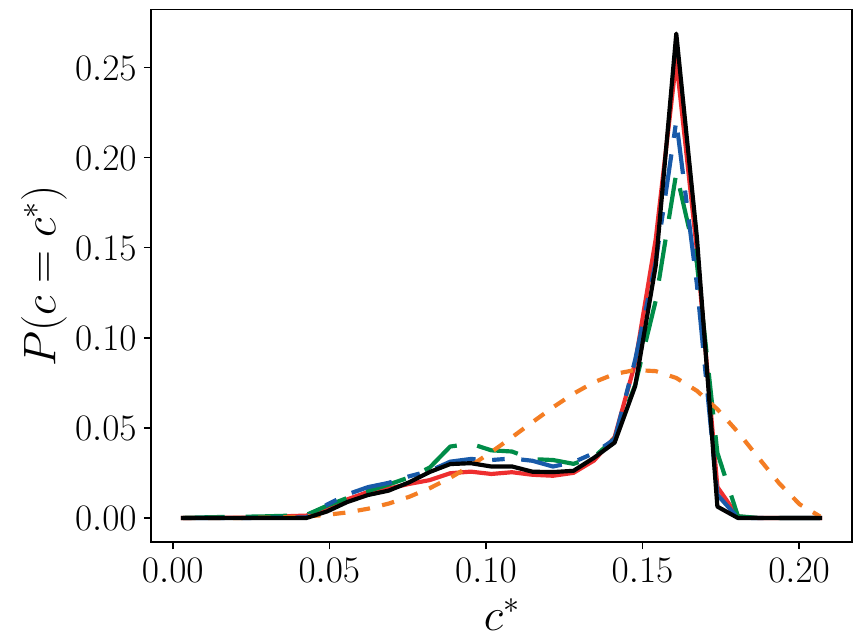}%
    \caption{$J(P||P_{\beta})=\nicefrac{\ln{(2)}}{2}$.}\label{fig:pdfs_hilo_2}%
  \end{subfigure}\hfill%
  \begin{subfigure}[t]{\textwidth}%
    \includegraphics[page=3,width=0.48\textwidth, trim=0.2cm 0cm 1.5cm 1.3cm, clip=true]{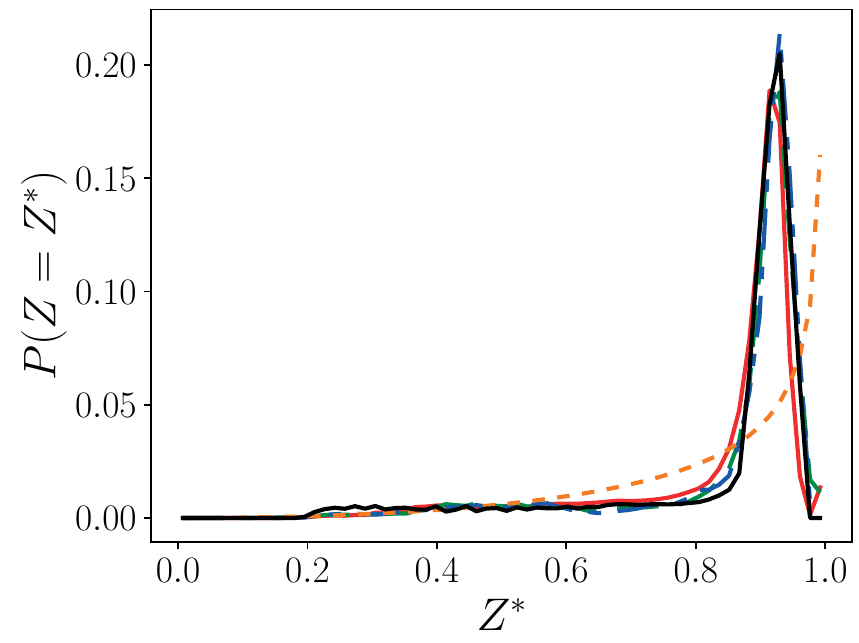}\hfill%
    \includegraphics[page=4,width=0.48\textwidth, trim=0.2cm 0cm 1.5cm 1.3cm, clip=true]{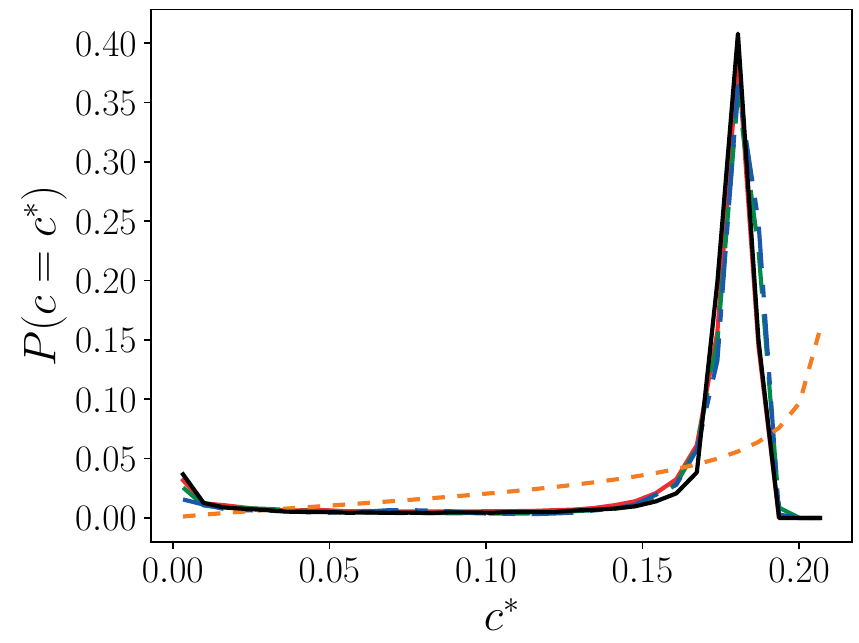}%
    \caption{$J(P||P_{\beta})=0.59$.}\label{fig:}%
  \end{subfigure}%
  \caption{Marginal \glspl{fdf} for low, mid-range, and high Jensen-Shannon divergence values for the $\beta$-$\beta$ \gls{pdf} model. Red solid: RF; green dashed: \gls{dnn}; blue dash-dotted: \gls{cvae}; orange short dashed: $\beta$-$\beta$ model; black solid: \gls{dns}.}\label{fig:pdfs_hilo}%
\end{figure}

We quantify model performance with two metrics of interest: the
Jensen-Shannon divergence~\cite{Endres2003, Osterreicher2003} and the
filtered progress variable source term. The Jensen-Shannon divergence
measures the similarity between two \glspl{pdf} and will characterize
the error in predicting $P(Z,c | \wt{Z}, \wt{Z''}, \wt{c},
\wt{c''})$. It is a symmetric version of the Kullback-Leibler
divergence~\cite{Kullback1987}, and it is defined as:
\begin{align}
  \label{eq:jsd}
  J(Q||R) = \frac{1}{2} \left( D(Q || M) + D(R || M)\right)
\end{align}
where
$D(Q||R) = \sum_{i=1}^n R(i) \ln{\left( \frac{R(i)}{Q(i)} \right)}$;
$M = \nicefrac{1}{2} \left( Q+R \right)$; $Q$ and $R$ are \glspl{pdf}
of length $n$; and $0\leq J(Q||R) \leq \ln{(2)}$, with low values
indicating more similarity between $Q$ and $R$. The Jensen-Shannon
divergence exhibits several advantages over the Kullback-Leibler
divergence: \glspl{pdf} do not need to have the same support, it is
symmetric, $J(Q||R) = J(R||Q)$, and it is bounded. The overall
sub-filter \gls{pdf} prediction accuracy of a model is characterized
by the $90^{\text{th}}$ percentile of all the Jensen-Shannon
divergences, denoted $J_{90}$. Examples of \gls{fdf}
modeling using the $\beta$-$\beta$ analytical model illustrate
different Jensen-Shannon divergence values,
Figure\,\ref{fig:pdfs_hilo}. This figure is similar to
Figure\,\ref{fig:pdfs}, though it shows different realizations of
$P(Z,c)$. The $\beta$-$\beta$ analytical model is not able to capture
more complex \gls{fdf} shapes, such as bimodal
distributions, leading to high Jensen-Shannon divergence values,
Figure\,\ref{fig:pdfs_hilo_2}, and it motivates the need for more
accurate models.  From these results, accurate predictions can be
expected for $J(P||P_m) < 0.3$, whereas predictions with
$J(P||P_m) > 0.6$ exhibit incorrect median values and overall shapes.

The second metric of interest characterizes the error in predicting
$\wt{\dot{\omega}}$ and is simply the normalized \gls{rmse} of the model predictions, $\wt{\dot{\omega}}_m$:
\begin{align}
  \label{eq:rmse}
  \text{RMSE}(\wt{\dot{\omega}}) = \frac{1}{\wt{\dot{\Omega}}}\sqrt{ \frac{1}{|\mathcal{D}|}\sum_{i=1}^{|\mathcal{D}|}\left( \epsilon(\wt{\dot{\omega}}_i) \right)^2},
\end{align}
where
$\epsilon(\wt{\dot{\omega}}_i) = \wt{\dot{\omega}}_i -
\wt{\dot{\omega}}_{m,i}$ is the error, $\mathcal{D}$ is the data set
over which the error is computed, and
\begin{align}
  \label{eq:norm}
  \wt{\dot{\Omega}} = \sqrt{\frac{1}{|\mathcal{D}_T|} \sum_{i=1}^{|\mathcal{D}_T|} \left( \wt{\dot{\omega}}_i \right)^2}
\end{align}
is the normalization constant, and
$\mathcal{D}_T = \bigcup\limits_{i=1, \dots, n_v} \mathcal{D}_i$. All metrics
presented are computed with respect to the validation data sets.

\subsection{\Acrlong{fdf} predictions}\label{sec:predictions}

\Gls{ml} models were trained using filtered \gls{dns} data from
$\mathcal{V}_3$ (centered at $z=0.1025\,\unit{m}$), i.e.,\ the
algorithms were trained on $\mathcal{D}_3^t$, and the metrics were
evaluated on $\mathcal{D}_3^v$. The random forests model training time
for the 52920 \glspl{fdf} in $\mathcal{D}_3^t$ was $1800\,\unit{s}$ on
an Intel SandyBridge Xeon processor with $256\,\unit{GB}$ of
memory. The \gls{dnn} and \gls{cvae} training times were
$2200\,\unit{s}$ and $3500\,\unit{s}$ on a NVIDIA Tesla K80 GPU. In
addition to the \gls{ml} models discussed in
Section\,\ref{sec:methods}, we included an \gls{ols} model as a
baseline for additional discussion of the modeling results.

\begin{figure}[!tbp]%
  \centering%
  \begin{subfigure}[t]{0.48\textwidth}%
    \includegraphics[page=1,width=\textwidth, trim=0.5cm 0cm 1.5cm 1.3cm, clip=true]{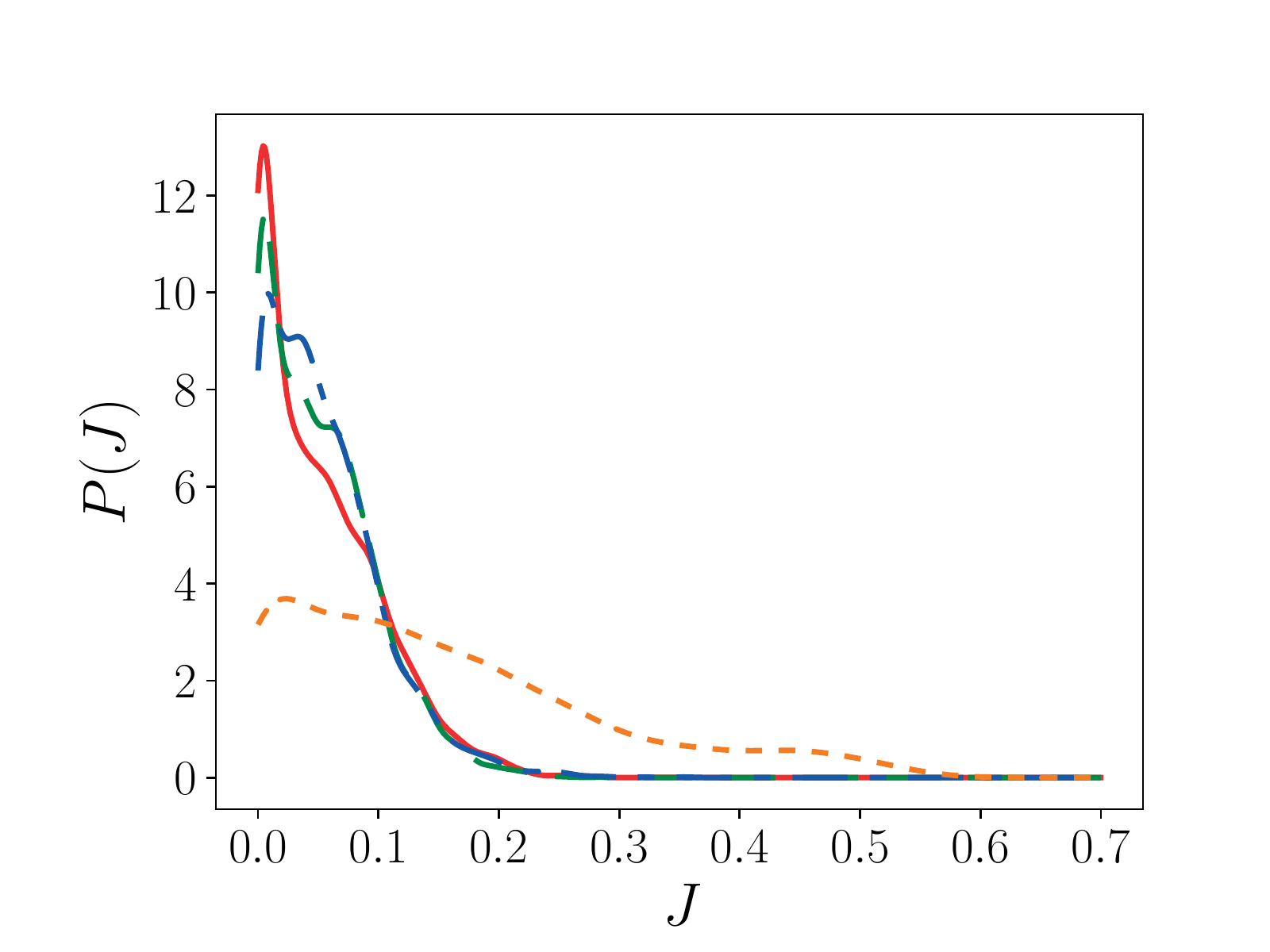}%
    \caption{\Gls{fdf} of $J$.}\label{fig:jsd_pdf}%
  \end{subfigure}\hfill%
  \begin{subfigure}[t]{0.48\textwidth}%
    \includegraphics[page=2,width=\textwidth, trim=0.5cm 0cm 1.5cm 1.3cm, clip=true]{./figs/jsd_dice_0004.pdf}%
    \caption{Cumulative density function of $J$.}\label{fig:jsd_cdf}%
  \end{subfigure}%
  \caption{\Gls{fdf} predictions on validation data. Red solid: RF; green dashed: \gls{dnn}; blue dash-dotted: \gls{cvae}; orange short dashed: $\beta$-$\beta$ model.}\label{fig:jsd}%
\end{figure}

Several example \gls{fdf} predictions are shown in
Figure\,\ref{fig:pdfs_hilo}, corresponding to low, medium, and high
values of $J(P||P_{\beta})$. The \gls{fdf} and the cumulative density
function for the Jensen-Shannon divergence of the predictions on the
validation data, $J=J(P||P_m)$, where $P_m$ is the modeled \gls{fdf},
are presented in Figure\,\ref{fig:jsd}. The three \gls{ml} models
exhibit similar \gls{fdf} prediction errors with a narrow peak close
to 0. The prediction error for the $\beta$-$\beta$ analytical model is
larger than the prediction errors for the \gls{ml} models; see
Table\,\ref{tab:summary}. Additionally, comparing the training error,
$J^t_{90}$, to the validation error, $J^v_{90}$, indicates that the
random forests model overfits the training data (there is a large
difference between the training and testing error), whereas the deep
learning algorithms avoid overfitting;
Table\,\ref{tab:summary}. Results from the \gls{ols} model, a low
complexity model, exhibit $J^t_{90}$ and $J^v_{90}$ values that are
approximately four times larger than that for the other \gls{ml}
models, Table\,\ref{tab:summary}. This indicates that the more complex
\gls{ml} models are capturing important complex physical effects that
the OLS approach misses.

Model prediction times for each \gls{fdf} were computed for all
models as:
\begin{align}
  \label{eq:predict}
  t_m = \frac{1}{n_t |\mathcal{D}_3^v|} \sum_{i=1}^{n_t} \text{time to evaluate } m(\mathcal{D}_3^v)
\end{align}
where $n_t=10$ predictions on the validation data set
$\mathcal{D}_3^v$, which contains 2940 samples, thereby necessitating
2940 model evaluations. Although the random forests model accuracy is
similar to that of the neural networks, the model complexity required
is such that the prediction time is approximately 20 times longer than
the \gls{dnn} and \gls{cvae} and the model size is more than 3000 times
larger; see Table\,\ref{tab:summary}. The need for large amounts of memory
for training and the slow prediction times illustrate the main
drawback for the use of the random forests algorithm in production
simulations from the standpoints of both training and prediction. Because the
\gls{dnn} and the \gls{cvae} decoder have similar architectures, their
prediction time is similar. The $\beta$-$\beta$ model 
\gls{fpdf} computations involve a discrete $\beta$ \gls{pdf} evaluation
in both $Z$ and $c$ and an outer product to compute the 
\gls{fpdf}, leading to prediction times comparable with the random
forests model. The $\beta$ \gls{pdf} was computed through the SciPy
library~\cite{Jones2001}.

\begin{table}[!tbp]
  \centering
  \begin{tabular}{lccccccc}
    \toprule
    Model & $J^t_{90}$ & $J^v_{90}$ & $t_m\,(\unit{ms})$ & \gls{rmse}$(\wt{\dot{\omega}})$& $R^2(\wt{\dot{\omega}})$ & \gls{dof} (million)& Memory (MB)\\
    \midrule
    RF            & 0.03 & 0.12 & 0.932 & 0.22 & 0.97 & 5.2 & 82107 \\
    \gls{dnn}     & 0.11 & 0.11 & 0.036 & 0.23 & 0.97 & 1.1 & 27 \\
    \gls{cvae}    & 0.11 & 0.12 & 0.038 & 0.22 & 0.97 & 2.3 & 36 \\
    $\beta$-$\beta$ & 0.35 & 0.35 & 1.178 & 0.63 & 0.75 & {--} & {--} \\
    \gls{ols}     & 0.43 & 0.43 & 0.018 & 0.69 & 0.70 & 0.008 & 0.1 \\
    \bottomrule
  \end{tabular}
  \caption{Summary of model performance and size for $P(Z,c | \wt{Z}, \wt{Z''}, \wt{c}, \wt{c''})$ and $\wt{\dot{\omega}}$.}\label{tab:summary}
\end{table}

\begin{figure}[!tbp]%
  \centering%
  \begin{subfigure}[t]{0.48\textwidth}%
    \includegraphics[page=1,width=\textwidth, trim=0.5cm 0cm 1.5cm 1.3cm, clip=true]{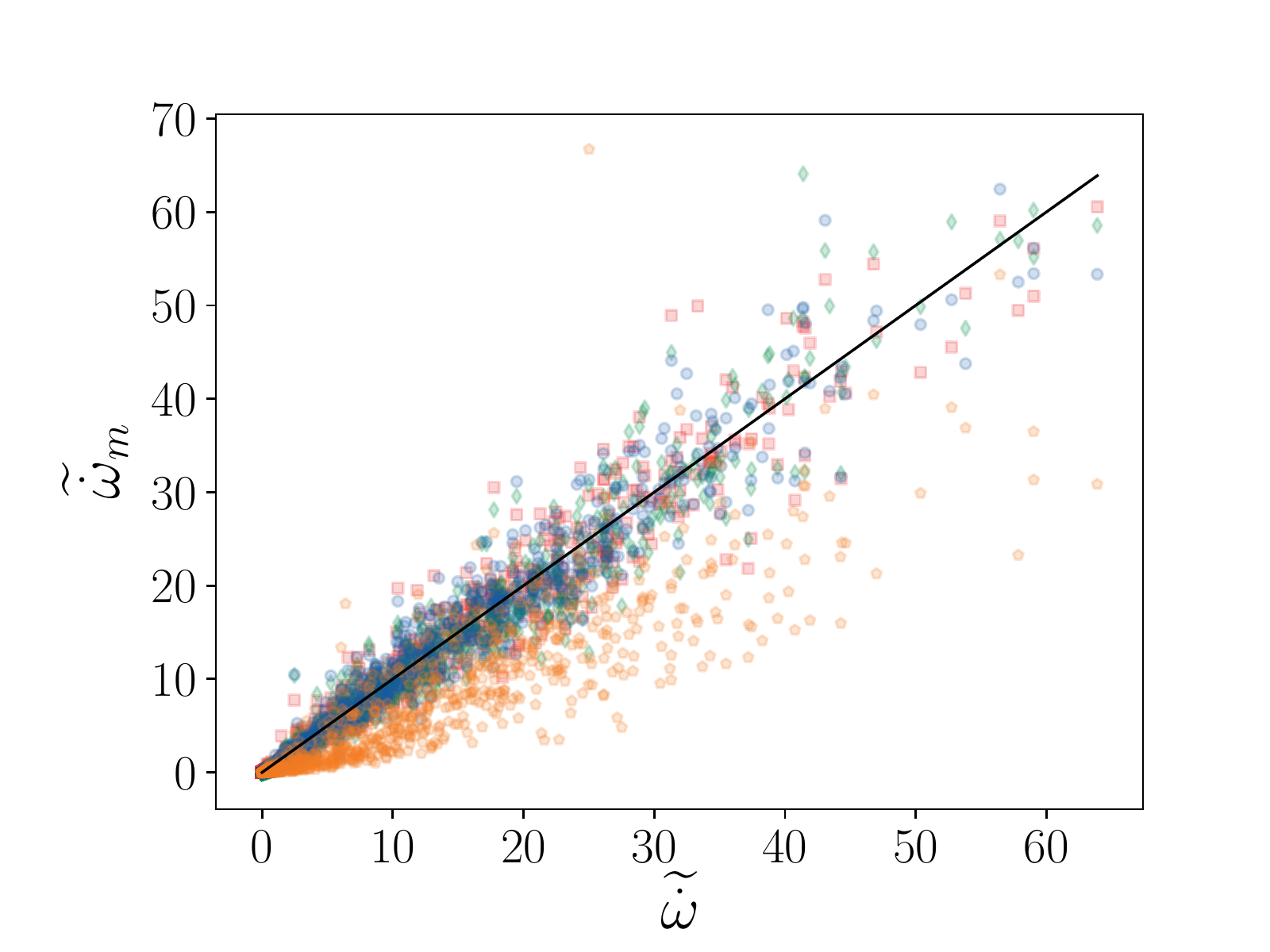}%
    \caption{$\wt{\dot{\omega}}$ predictions.}\label{fig:convolution_scatter}%
  \end{subfigure}\hfill%
  \begin{subfigure}[t]{0.48\textwidth}%
    \includegraphics[page=2,width=\textwidth, trim=0.5cm 0cm 1.5cm 1.3cm, clip=true]{./figs/convolution_dice_0004.pdf}%
    \caption{\gls{pdf} of $\epsilon(\wt{\dot{\omega}})$.}\label{fig:convolution_pdf}%
  \end{subfigure}%
  \caption{Reaction rate predictions. Red squares and solid: RF; green diamonds and dashed: \gls{dnn}; blue circles and dash-dotted: \gls{cvae}; orange pentagons and short dashed: $\beta$-$\beta$ model.}\label{fig:convolution}%
\end{figure}

The \gls{fdf} models were used to provide predictions of the reaction
rate, $\wt{\dot{\omega}}$, by convoluting the predicted \gls{fdf} with
the reaction rate, Equation\,\ref{eq:convolution}, where
$\langle \dot{\omega} | Z, c \rangle$ is from the same $32^3$ box as
that used to generate $P(Z,c | \wt{Z}, \wt{Z''}, \wt{c},
\wt{c''})$. This ensures that the errors observed in the predictions
of $\wt{\dot{\omega}}$ can be exclusively attributed to the \gls{fdf}
modeling. Table\,\ref{tab:summary} and Figure\,\ref{fig:convolution}
illustrate the different model performances in predicting
$\wt{\dot{\omega}}$. The coefficient of determination, $R^2$, is above
$0.95$ for the three discussed \gls{ml} models, indicating a high
model accuracy, whereas that of the $\beta$-$\beta$ model is
significantly lower. The three different \gls{ml} algorithms achieve
similar results, Figure\,\ref{fig:convolution}. The \gls{pdf} of the
error, $\epsilon(\wt{\dot{\omega}})$, as shown in
Figure\,\ref{fig:convolution_pdf}, is symmetric, indicating that the
models are not biased toward under- or overpredicting. The
$\beta$-$\beta$ analytical model has a broad range of prediction
errors and tends to underpredict $\wt{\dot{\omega}}$ for
$\wt{\dot{\omega}} > 5$ in this volume,
Figure\,\ref{fig:convolution_scatter}. The \gls{ols} model presents an
error that is three times larger than the other \gls{ml} models and is
slightly larger than the error of the $\beta$-$\beta$ model.

\subsection{Model generalization}\label{sec:predictions_all}

\begin{figure}[!tbp]%
  \centering%
  \includegraphics[page=4,width=0.6\textwidth]{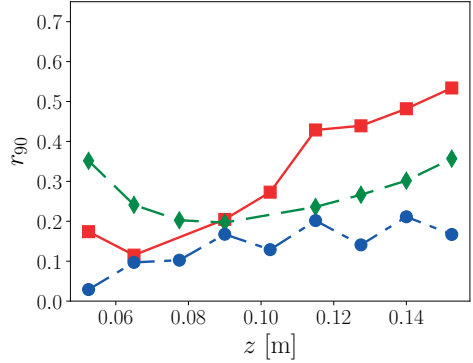}\\%
  \caption{The $90^{\text{th}}$ percentile of $r_{i,j}$ as a function
    of height. Red squares and solid: $r_{3,j}$ for $j=1,\dots,n_v$;
    green diamonds and dashed: $r_{5,j}$ for $j=1,\dots,n_v$; blue
    circles and dash-dotted: $r_{i,j}$ for $i=[1,3,5,7,9]$ and
    $j=2,4,6,\,\text{and}\ 8$.}\label{fig:r90}%
\end{figure}%

In this section, we examine the performance of the models trained
using different data sets and their ability to generalize to data from
other regions of the flame. An understanding of model generalization
has important implications for the model's applicability to other
physical configurations. Because of the nature of the low-swirl burner
flame, a wide range of different physical processes are encountered in
different regions of the flame, and it is important to understand the
conditions for a model's applicability.

Three versions of the models were trained: (i) using $\mathcal{D}_3^t$
(volume centered at $z=0.0775\,\unit{m}$); (ii) using data from a volume
farther downstream, $\mathcal{D}_5^t$ (volume centered at
$z=0.1025\,\unit{m}$); and (iii) trained using data from every other
volume
$\mathcal{D}^t = \bigcup\limits_{i=1, 3, 5, 7, 9} \mathcal{D}_i^t$.
Note that for the random forests trained on all volumes, the maximum depth size
of the trees was reduced to 18 to avoid out-of-memory errors
on a 256 GB node (the resulting model size exceeded 110 GB).

The difference between the \glspl{fdf} in different volumes is quantified
through the minimum of the pairwise Jensen-Shannon divergence between
all \glspl{fdf} belonging to $\mathcal{V}_i$ and all \glspl{fdf} belonging to
$\mathcal{V}_j$:
\begin{align}
  \label{eq:distance}
  r_{i,j} = \min_{k=1,\dots, |\mathcal{D}_i^v|} J( P_k || P_l )\quad \forall P_k \in \mathcal{D}_i^v,\ \forall P_l \in \mathcal{D}_j^v.
\end{align}
Low values of $r_{i,j}$ indicate that
$\forall P_l \in \mathcal{D}_j^v$ there is $P_k \in \mathcal{D}_i^v$,
which has a small Jensen-Shannon divergence and, therefore, a similar
shape. The $90^{\text{th}}$ percentile of $r_{i,j}$, $r_{90}$, for
different data sets is presented in Figure\,\ref{fig:r90}. For
$\mathcal{V}_3$, it is clear that the \glspl{fdf} in regions of the flame
that are farther downstream or upstream are significantly
different; however, models trained using data from every other volume,
$\mathcal{D}^t$, have training data that are representative of the
entire simulation domain.

\begin{figure}[!tbp]%
  \centering%
  \begin{subfigure}[t]{\textwidth}%
    \includegraphics[page=1,width=0.48\textwidth, trim=0.2cm 0cm 1cm 1cm, clip=true]{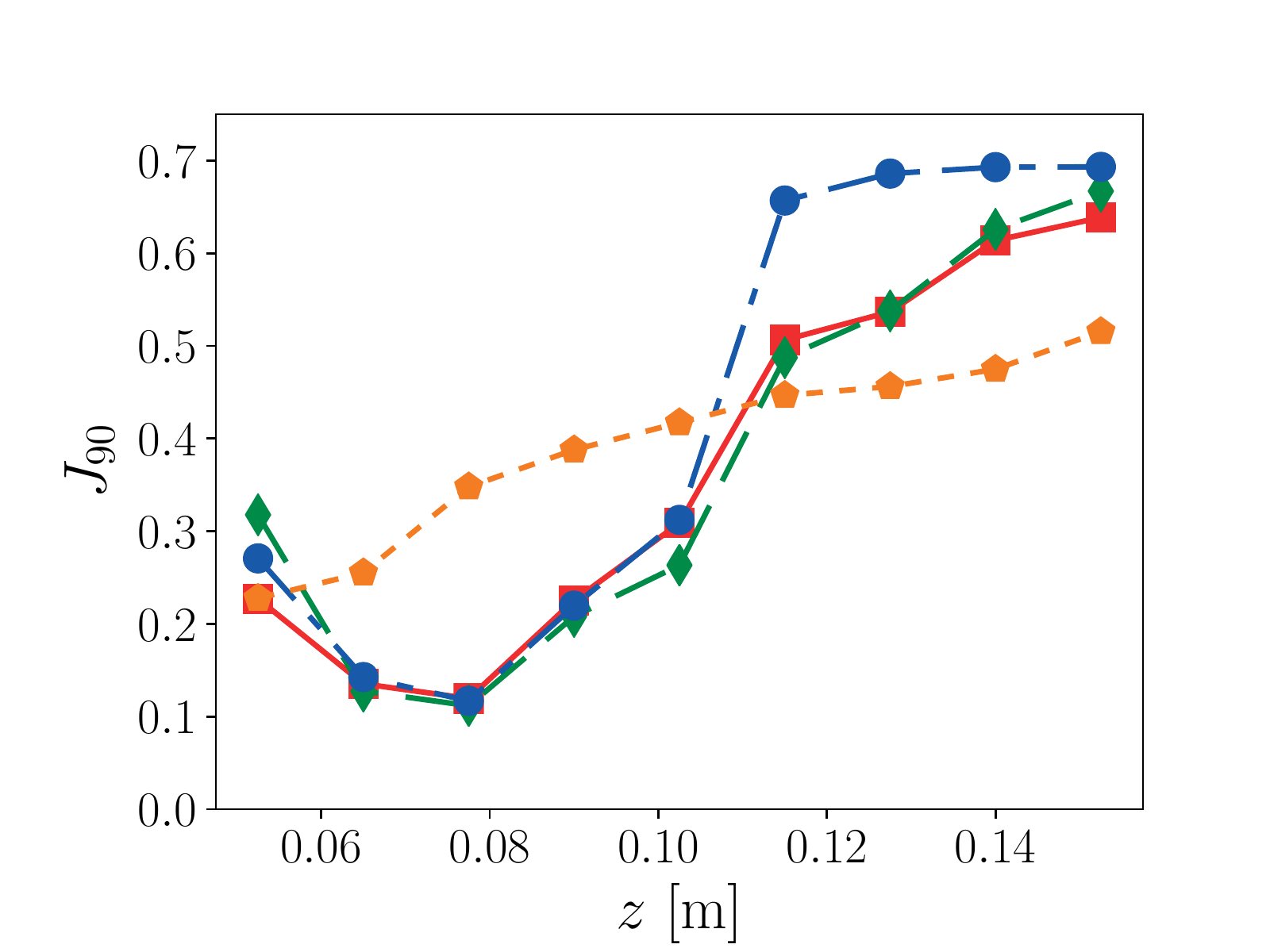}\hfill%
    \includegraphics[page=2,width=0.48\textwidth, trim=0.2cm 0cm 1cm 1cm, clip=true]{./figs/dice_predictions_0004.pdf}%
    \caption{$\mathcal{D}_3^t$ ($z_3=0.0775\,\unit{m}$).}\label{fig:gen_3}%
  \end{subfigure}\hfill%
  \begin{subfigure}[t]{\textwidth}%
    \includegraphics[page=1,width=0.48\textwidth, trim=0.2cm 0cm 1cm 1cm, clip=true]{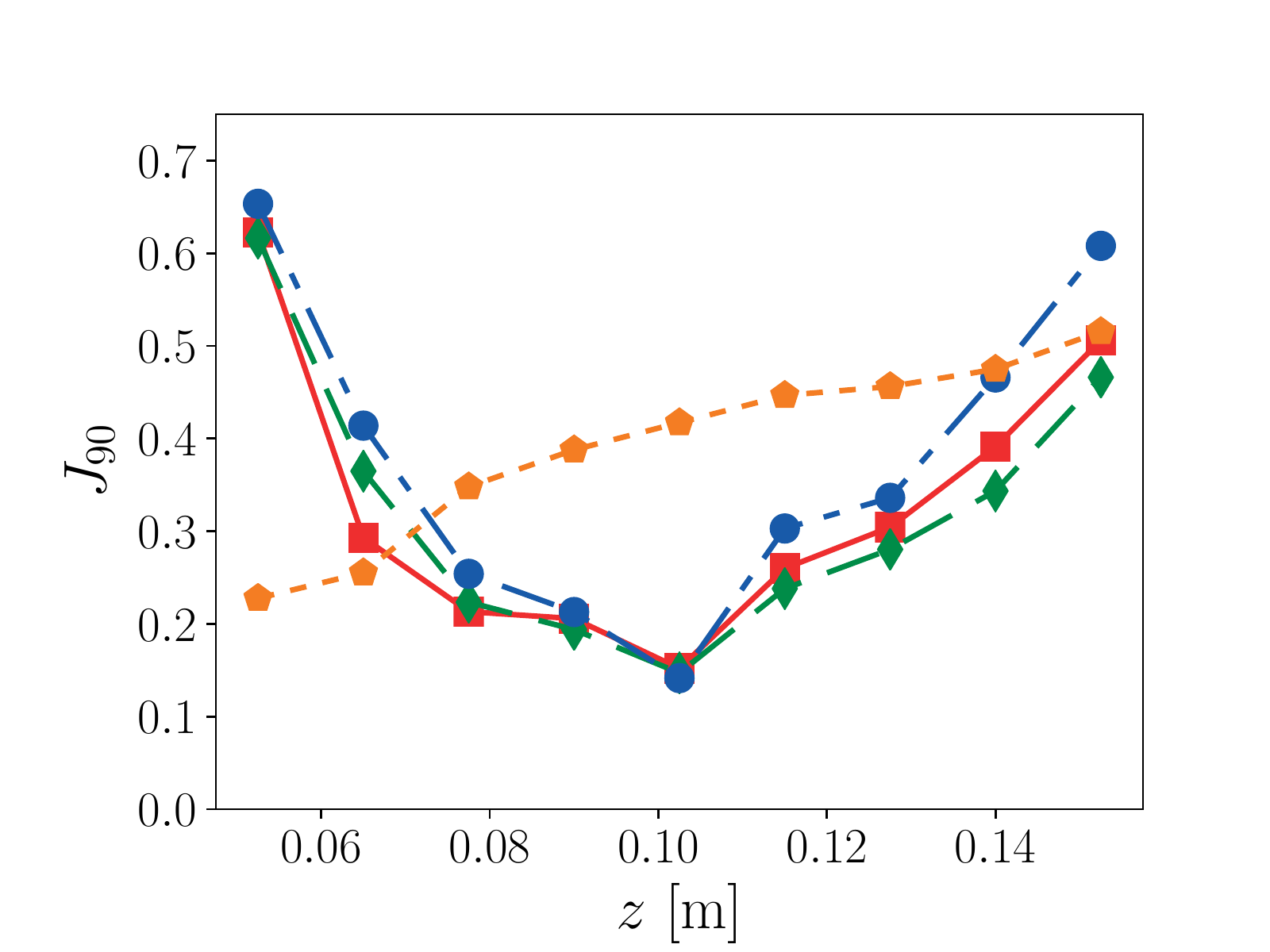}\hfill%
    \includegraphics[page=2,width=0.48\textwidth, trim=0.2cm 0cm 1cm 1cm, clip=true]{./figs/dice_predictions_0006.pdf}%
    \caption{$\mathcal{D}_5^t$ ($z_5=0.1025\,\unit{m}$).}\label{fig:gen_5}%
  \end{subfigure}\hfill%
  \begin{subfigure}[t]{\textwidth}%
    \includegraphics[page=1,width=0.48\textwidth, trim=0.2cm 0cm 1cm 1cm, clip=true]{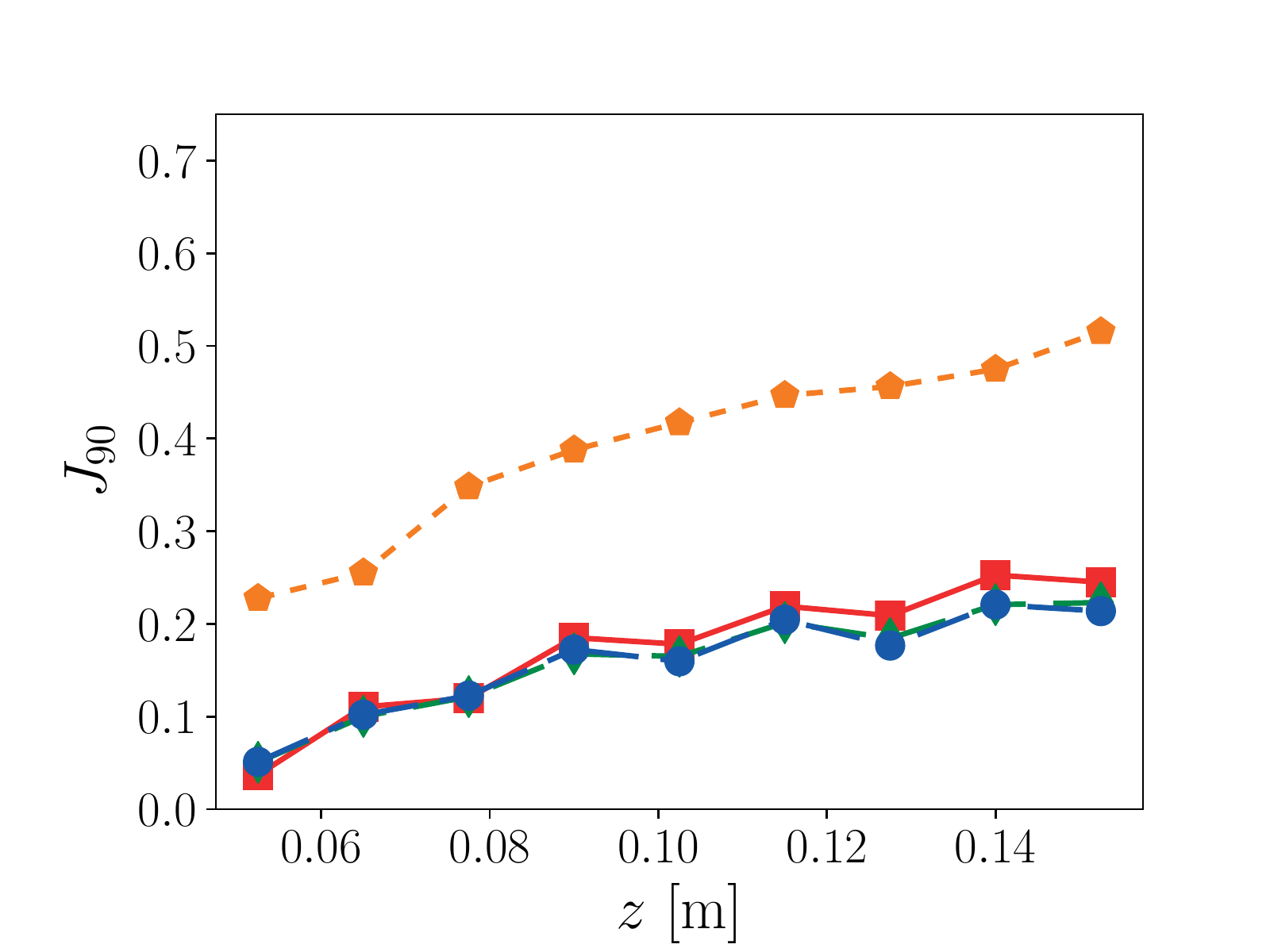}\hfill%
    \includegraphics[page=2,width=0.48\textwidth, trim=0.2cm 0cm 1cm 1cm, clip=true]{./figs/dice_predictions_skip.pdf}%
    \caption{$\mathcal{D}^t = \bigcup\limits_{i=1, 3, 5, 7, 9} \mathcal{D}_i^t$.}\label{fig:gen_skip}%
  \end{subfigure}%
  \caption{$J_{90}$ and \gls{rmse}$(\wt{\dot{\omega}})$ as a function of
    height using \gls{ml} algorithms trained with data from different
    sections of the flame. Red squares and solid: RF; green diamonds
    and dashed: \gls{dnn}; blue circles and dash-dotted: \gls{cvae}; orange
    pentagons and short dashed: $\beta$-$\beta$ model.}\label{fig:gen}%
\end{figure}

Figure\,\ref{fig:gen} presents the predictions for the three different
model versions. For models trained using data from only one volume,
the \gls{fdf} prediction error is lowest for that volume and increases as
the model is used on downstream or upstream volumes.  All three types
of \gls{ml} algorithms predict similar generalization error
profiles. This indicates that these models, including the generative
algorithm, are unable to extrapolate to non-proximate regions of the
flame. This is consistent with the observation that the training data
are not representative of the entire flow, Figure\,\ref{fig:r90}. The
\gls{rmse}$(\wt{\dot{\omega}})$ decreases as a function of $z$ because the
mean $\wt{\dot{\omega}}$ decreases as a function of $z$ as
well. Models trained with $\mathcal{D}_5^t$ perform slightly better in
the upstream portion of the domain, Figure\,\ref{fig:gen_5}, because
the \glspl{fdf} in $\mathcal{V}_5$ are more representative of the upstream
\glspl{fdf}, but fail to capture those where the premixed burning at the
nozzle is dominant ($z\approx 0.05\,\unit{m}$), Figure\,\ref{fig:r90}.

\begin{figure}[!tbp]%
  \centering%
  \begin{subfigure}[t]{0.48\textwidth}%
    \includegraphics[page=1,width=\textwidth, trim=0.5cm 0cm 1.3cm 1.3cm, clip=true]{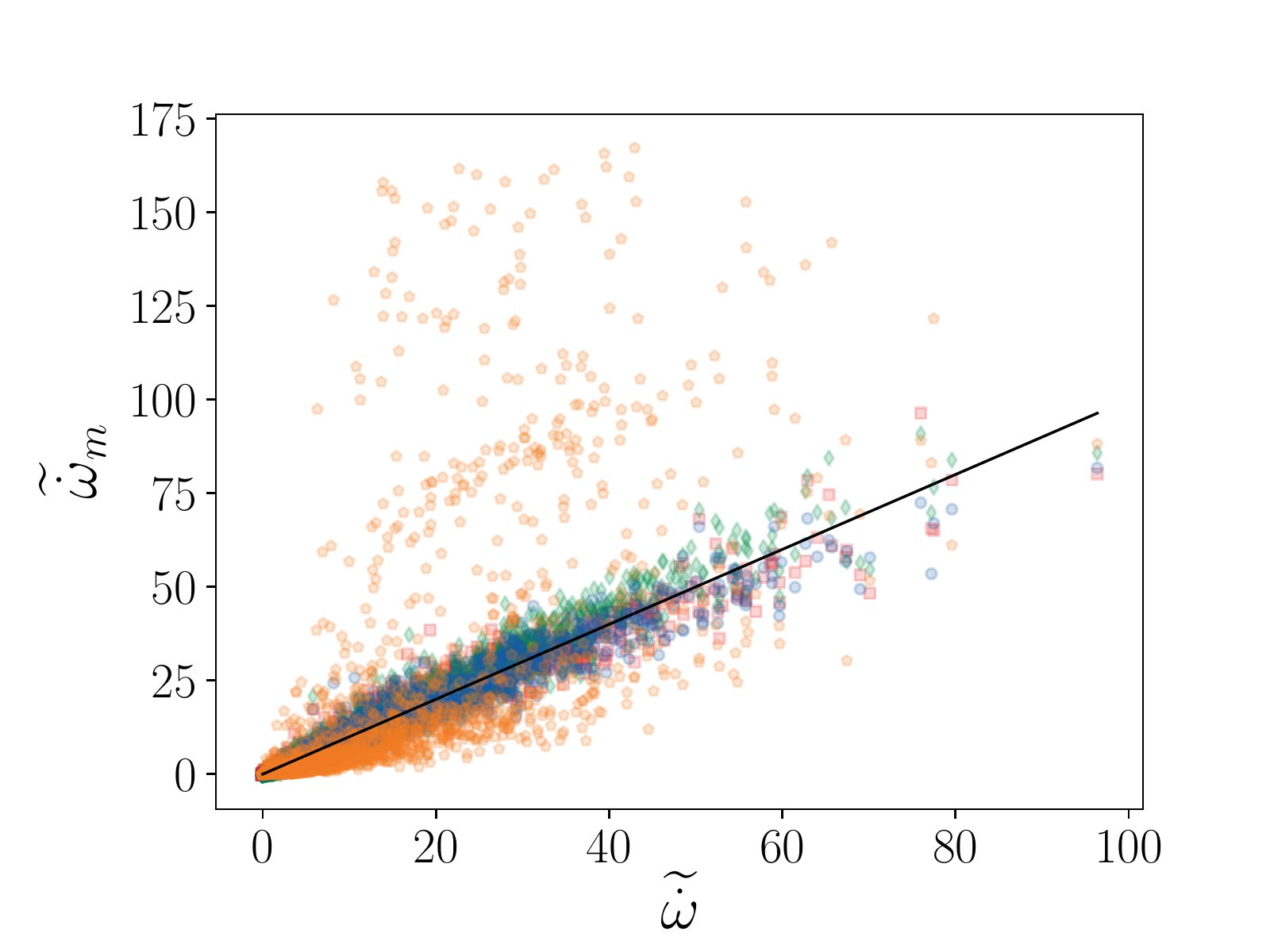}%
    \caption{$\wt{\dot{\omega}}$ predictions.}%
  \end{subfigure}\hfill%
  \begin{subfigure}[t]{0.48\textwidth}%
    \includegraphics[page=3,width=\textwidth, trim=0.5cm 0cm 1.5cm 1.3cm, clip=true]{./figs/convolution_skip.pdf}%
    \caption{\mbox{Conditional means of $\wt{\dot{\omega}}$ as a function of $\wt{z}$.}}%
  \end{subfigure}\\[0.2cm]%
  \begin{subfigure}[t]{0.48\textwidth}%
    \includegraphics[page=4,width=\textwidth, trim=0.5cm 0cm 1.5cm 1.3cm, clip=true]{./figs/convolution_skip.pdf}%
    \caption{\mbox{Conditional means of $\wt{\dot{\omega}}$ as a function of $\wt{c}$.}}%
  \end{subfigure}%
  \caption{Reaction rate predictions for models trained on $\mathcal{D}^t = \bigcup\limits_{i=1, 3, 5, 7, 9} \mathcal{D}_i^t$. Red squares and solid: RF; green diamonds and dashed: \gls{dnn}; blue circles and dash-dotted: \gls{cvae}; orange pentagons and short dashed: $\beta$-$\beta$ model; black solid: \gls{dns}.}\label{fig:convolution_skip}%
\end{figure}

\begin{figure}[!tbp]%
  \centering%
  \begin{subfigure}[t]{\textwidth}%
    \includegraphics[page=3,width=0.48\textwidth, trim=0.2cm 0cm 1cm 1cm, clip=true]{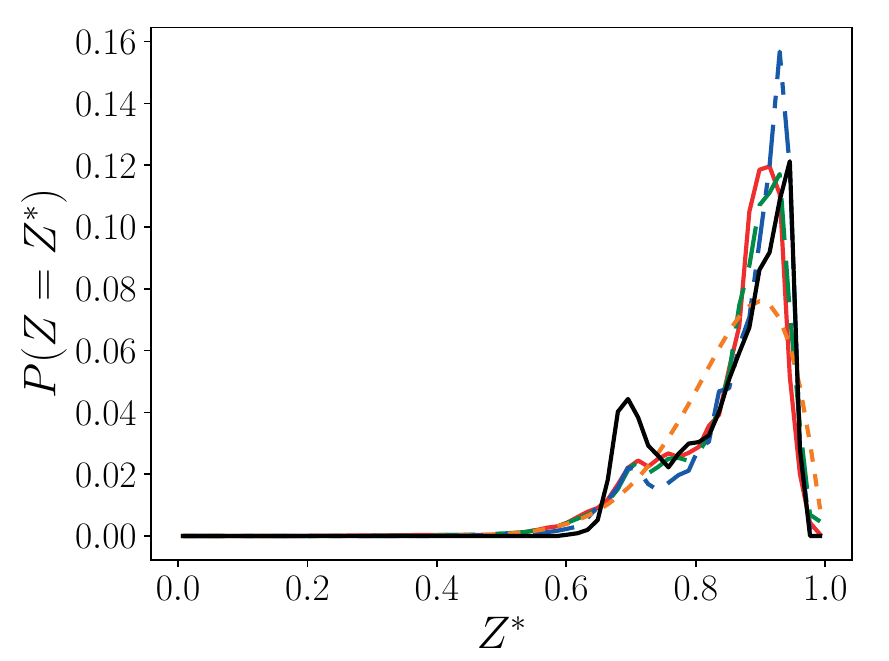}\hfill%
    \includegraphics[page=4,width=0.48\textwidth, trim=0.2cm 0cm 1cm 1cm, clip=true]{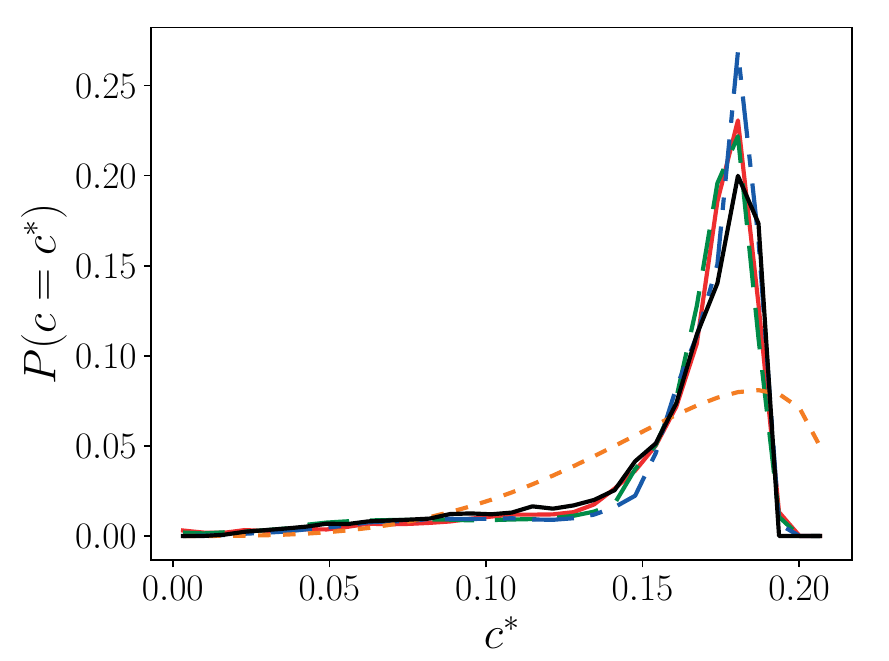}%
    \caption{$J(P||P_{\text{\gls{dnn}}})=0.05$.}\label{fig:}%
  \end{subfigure}\hfill%
  \begin{subfigure}[t]{\textwidth}%
    \includegraphics[page=3,width=0.48\textwidth, trim=0.2cm 0cm 1cm 1cm, clip=true]{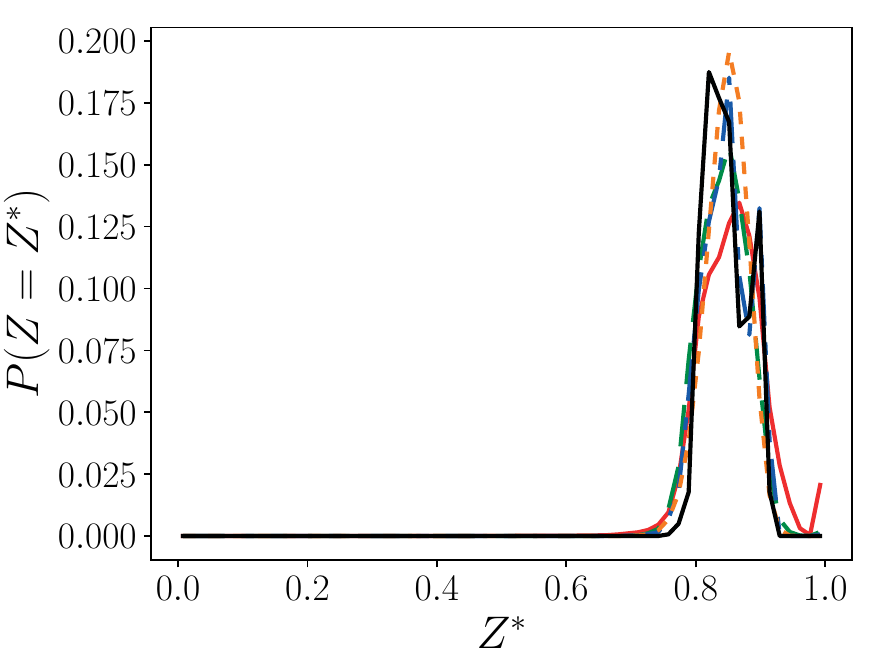}\hfill%
    \includegraphics[page=4,width=0.48\textwidth, trim=0.2cm 0cm 1cm 1cm, clip=true]{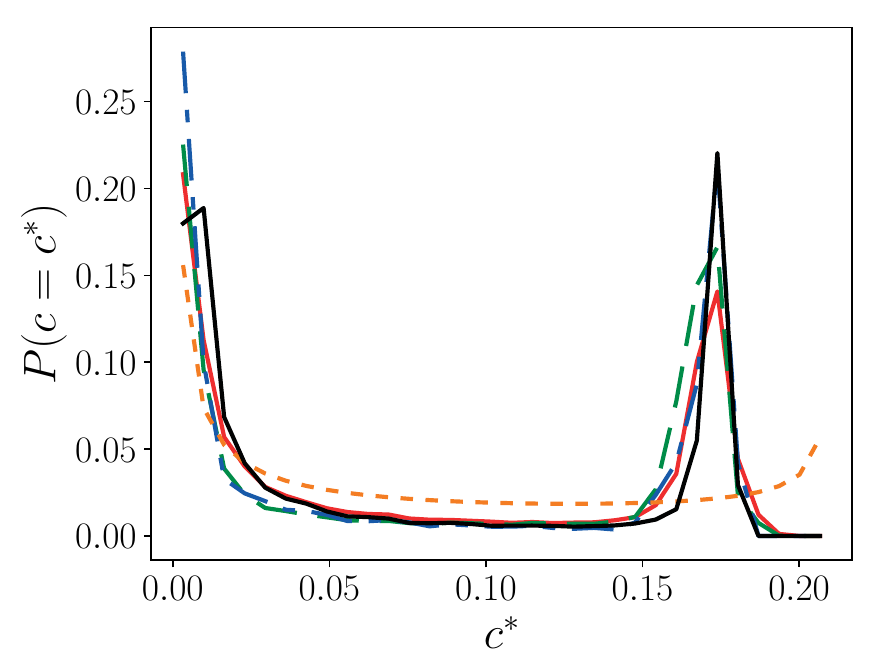}%
    \caption{$J(P||P_{\text{\gls{dnn}}})=0.1$.}\label{fig:}%
  \end{subfigure}\hfill%
  \begin{subfigure}[t]{\textwidth}%
    \includegraphics[page=3,width=0.48\textwidth, trim=0.2cm 0cm 1cm 1cm, clip=true]{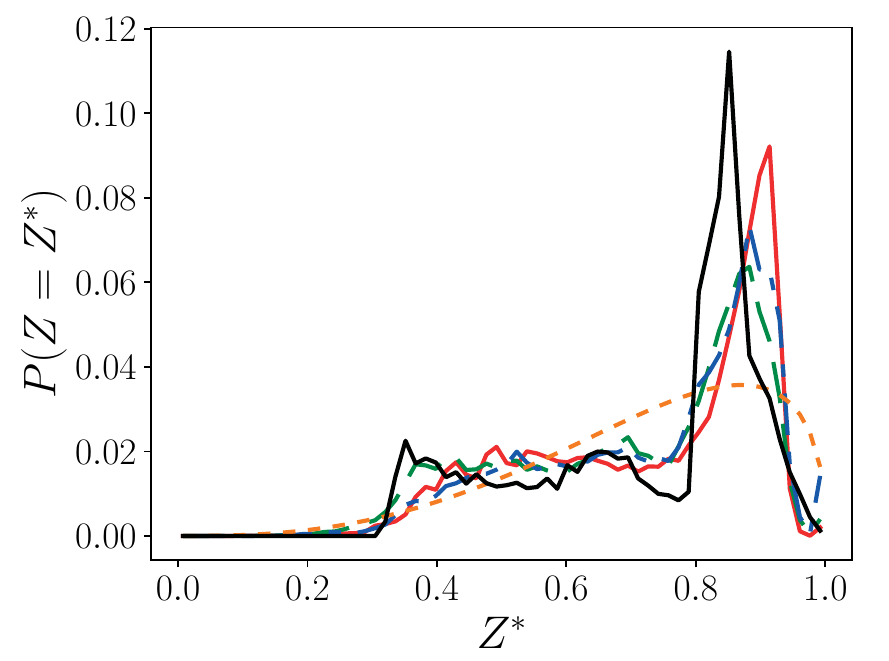}\hfill%
    \includegraphics[page=4,width=0.48\textwidth, trim=0.2cm 0cm 1cm 1cm, clip=true]{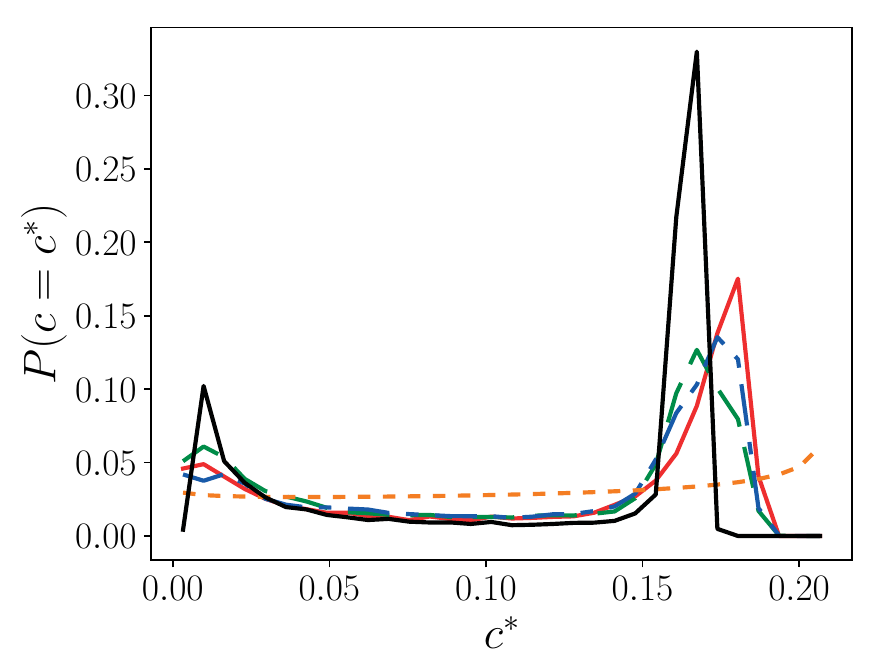}%
    \caption{$J(P||P_{\text{\gls{dnn}}})=0.21$.}\label{fig:pdfs_skip_3}%
  \end{subfigure}%
  \caption{Marginal \glspl{fdf} for median and high Jensen-Shannon divergence values for models trained on $\mathcal{D}^t = \bigcup\limits_{i=1, 3, 5, 7, 9} \mathcal{D}_i^t$. Red solid: RF; green dashed: \gls{dnn}; blue dash-dotted: \gls{cvae}; orange short dashed: $\beta$-$\beta$ model; black solid: \gls{dns}.}\label{fig:pdfs_skip}%
\end{figure}

Models trained using every other volume achieve errors that are
approximately half the error of the $\beta$-$\beta$ analytical model,
Figure\,\ref{fig:gen_skip}. This indicates that the models are capable
of interpolating the sample space across the entire physical domain
while using only a small subsection of the samples in the domain. The
\gls{ml} models achieve very good accuracy and approximate the
conditional means of $\wt{\dot{\omega}}$, which is the optimal
estimator using these data,
Figure\,\ref{fig:convolution_skip}. Significant overpredictions in
the $\beta$-$\beta$ model are observed. These are driven by errors in
upstream volumes, Figure\,\ref{fig:gen_skip}, particularly at high
$\wt{Z}$ and $\wt{c}$, Figure\,\ref{fig:convolution_skip}. Sample \glspl{fdf}
where $\wt{\dot{\omega}} > 15$ are shown in
Figure\,\ref{fig:pdfs_skip} for different Jensen-Shannon divergences
computed on the \gls{dnn} model. Bimodal distributions are accurately
predicted by the \gls{ml} models, and, even for the worse
case, Figure\,\ref{fig:pdfs_skip_3}, the shapes in $Z$ and $c$ are well
modeled. 

In addition to demonstrating the accuracy of the \gls{ml}
algorithms, these results illustrate that the $90^{\text{th}}$
percentile of $r_{i,j}$ is a good metric for characterizing \gls{fdf}
similarity and provides a model generalization criteria,
$r_{90} < 0.2$, for an a-priori assessment of model performance on new
data. Models trained using a data set that has an $r_{90} < 0.2$ with
another data set will produce joint \glspl{fdf} exhibiting
$J_{90} < 0.2$ and, consequently, accurate $\wt{\dot{\omega}}$
predictions. As a demonstration, a \gls{dnn} model was trained using
samples from the negative $x$-half of the volume $\mathcal{V}_3$ of
the axisymmetric flame (centered at $x=y=0\,\unit{m}$), i.e.,\
$\mathcal{D}^t = \{ s~|~s \in \mathcal{D}_3^t,~x_s < 0\,\unit{m}\}$, and
validated on predictions of samples in the positive $x$-half of the
other volumes,
$\mathcal{D}^v = \{ s~|~s \in \mathcal{D}_i^v,~x_s >
0\,\unit{m},~i=1,\dots, n_v\}$. This model performs accurately on
\gls{fdf} predictions in nearby volumes, e.g., $J_{90} \approx 0.15$ in
$\mathcal{V}_2$ and $\mathcal{V}_3$, and performs poorly at locations
farthest downstream, e.g.,\ $J_{90} = 0.63$ in $\mathcal{V}_7$.

\begin{figure}[!tbp]%
  \centering%
  \begin{subfigure}[t]{0.48\textwidth}%
    \includegraphics[page=1,width=\textwidth, trim=0.5cm 0cm 1.3cm 1.3cm, clip=true]{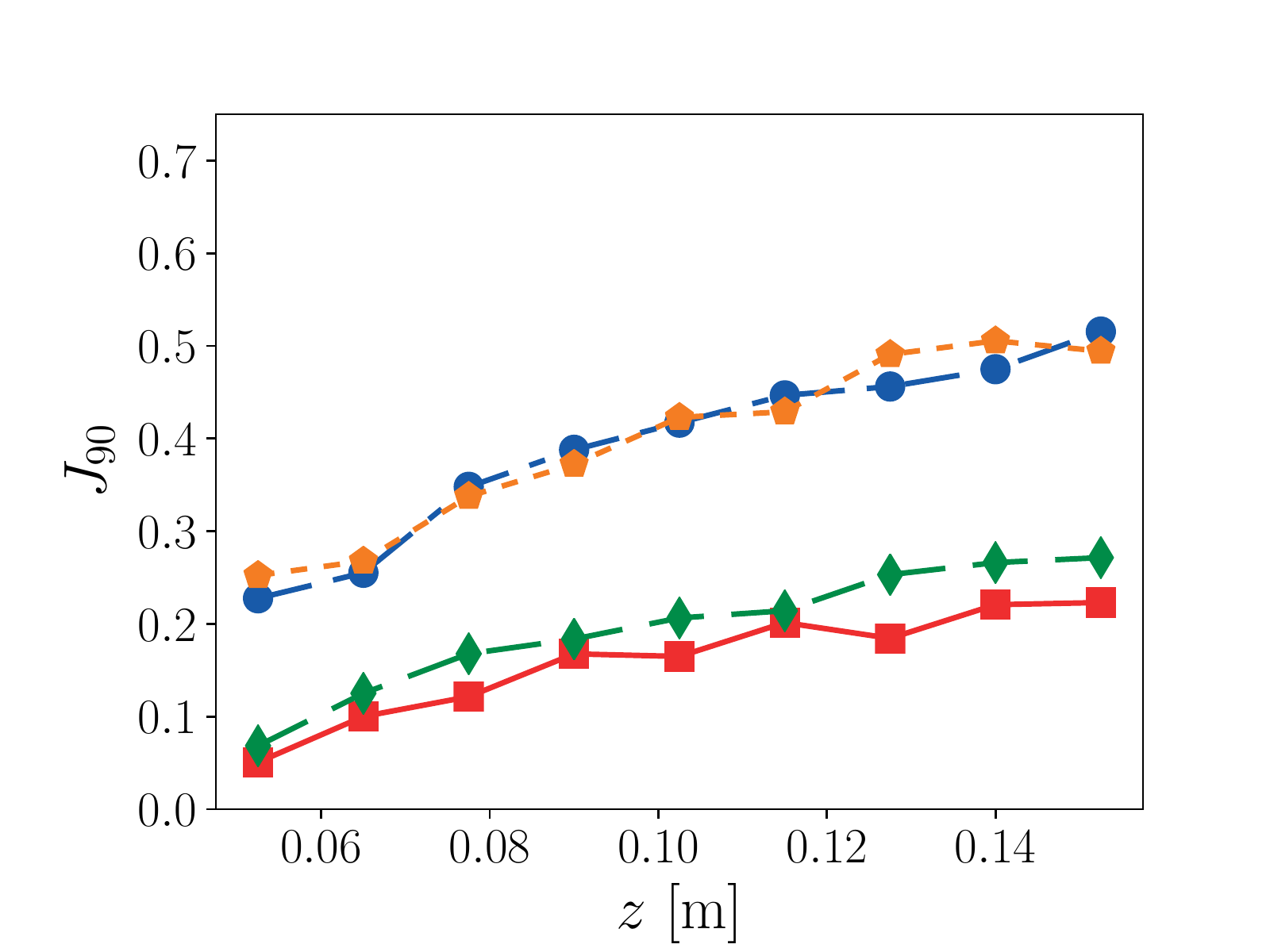}%
    \caption{$J_{90}$.}%
  \end{subfigure}\hfill%
  \begin{subfigure}[t]{0.48\textwidth}%
    \includegraphics[page=2,width=\textwidth, trim=0.5cm 0cm 1.5cm 1.3cm, clip=true]{./figs/dice_predictions_comp_1030.pdf}%
    \caption{\gls{rmse}$(\wt{\dot{\omega}})$.}%
  \end{subfigure}%
  \caption{$J_{90}$ and \gls{rmse}$(\wt{\dot{\omega}})$ as a function
    of height using the \gls{dnn} model trained at
    $t=0.0626\,\unit{s}$ on
    $\mathcal{D}^t = \bigcup\limits_{i=1, 3, 5, 7, 9} \mathcal{D}_i^t$
    and the $\beta$-$\beta$ model. Red squares and solid: \gls{dnn} at
    $t=0.0626\,\unit{s}$; green diamonds and dashed: \gls{dnn} at
    $t=0.059\,\unit{s}$; blue circles and dash-dotted:
    $\beta$-$\beta$ model at $t=0.0626\,\unit{s}$; orange pentagons
    and short dashed: $\beta$-$\beta$ model at
    $t=0.059\,\unit{s}$.}\label{fig:prediction_comp_1030}%
\end{figure}

The last generalization test was performed by using data generated
from a different time snapshot of the \gls{dns} ($t=0.059\,\unit{s}$)
as that used to train the models. For this case, the \gls{dnn} model
trained on
$\mathcal{D}^t = \bigcup\limits_{i=1, 3, 5, 7, 9} \mathcal{D}_i^t$ and
the $\beta$-$\beta$ model were used to illustrate how these models
perform on data from a different time instance of the \gls{dns},
Figure\,\ref{fig:prediction_comp_1030}. The \gls{dnn} model predicts
similar $J_{90}$ values though they are slightly higher for the data
from the time snapshot not used in the training. The $\beta$-$\beta$
model presents similar errors in both cases and these remain
approximately three times higher than those of the \gls{dnn}
model. These generalization tests clearly demonstrate that the
learned models are able to generalize temporally, as well as
spatially.

The results in this section illustrate (i) the importance of using
data representative of the extent of the physical processes present in
the simulation and (ii) the potential to develop in situ \gls{ml}
modeling capabilities, where the model is developed during the
simulation, without adversely affecting the simulation time because
the most accurate models were trained using data from less than $4\%$
of the total \gls{dns} domain volume.

\section{Conclusion}\label{sec:ccl}

In this work, we used three different \gls{ml} algorithms
representative of different types of \gls{ml} (traditional methods,
deep learning, and generative learning) to design presumed \gls{pdf}
models for combustion applications. We showed that models designed
through advanced \gls{ml} techniques are better able to capture the
complexity of these \glspl{fdf} than analytical models or linear
regression models. Although the random forests model predicts results
similar to those of the deep learning models, this model is not
suitable for in situ training and modeling because of the model
complexity, which leads to high memory requirements and high
prediction times. The deep learning algorithms were able to achieve
the same high accuracy with fast prediction times and low model
complexity. These models were also able to generalize to other spatial
regions of the flame and to other time instances of the flame.  In
this study, generative learning models as used here, which present
advantages in many deep learning applications through the use of a
latent space representation, did not provide increased accuracy or
better generalization characteristics compared to feed-forward neural
networks. Additionally, the deep learning models provide fast
predictions relative to the $\beta$-$\beta$ model, indicating that
these methods might at the very least be efficient encoders of
$\beta$-$\beta$ tabulation models by using \gls{dns} as a source of
training data, resulting in encodings that provide more useful forms
of the joint \gls{fdf} not expressible by the $\beta$-$\beta$
model. Our results illustrate methodologies that can be successfully
leveraged to derive accurate deep learning models for a wide range of
applications. The results exhibited throughout this work indicate that
deep learning models can be advantageously used for in situ modeling
of turbulent combustion flows. These deep learning algorithms are
readily integrated with scientific computing codes through PyTorch's
C\texttt{++} API for future a-posteriori model evaluation. This work
explores the construction of presumed \gls{pdf} models parameterized
by the commonly used four moments of the joint \gls{fdf} using
\gls{ml} techniques. The existence of a unique \gls{fdf}
representation would require the specification of all the \gls{fdf}
moments. The lack of uniqueness is demonstrated in the results
presented in this paper where a model constructed with a certain
subset of the data failed to generalize to other regions of the
flame. Addressing the non-uniqueness of the joint \glspl{fpdf} will
require the combustion community to explore modeling paradigms that do
not solely rely on a finite set of \gls{fdf} moments.

This work --- including neural network models, analysis scripts, Jupyter
notebooks, and figures --- can be publicly accessed at the project's
GitHub
page.\footnote{\url{https://github.com/NREL/ml-combustion-pdf-models}}
Traditional \gls{ml} algorithms were implemented through
scikit-learn~\cite{Pedregosa2011} and the deep learning algorithms
through PyTorch~\cite{Paszke2017}.

\section*{Acknowledgments}
This work was authored in part by the National Renewable Energy Laboratory, operated by Alliance for Sustainable Energy, LLC, for the U.S. Department of Energy (DOE) under Contract No. DE-AC36-08GO28308. Funding provided by U.S. Department of Energy Office of Science and National Nuclear Security Administration. The views expressed in the article do not necessarily represent the views of the DOE or the U.S. Government. The U.S. Government retains and the publisher, by accepting the article for publication, acknowledges that the U.S. Government retains a nonexclusive, paid-up, irrevocable, worldwide license to publish or reproduce the published form of this work, or allow others to do so, for U.S. Government purposes.

This research was supported by the Exascale Computing Project (ECP), Project Number: 17-SC-20-SC, a collaborative effort of two DOE organizations -- the Office of Science and the National Nuclear Security Administration -- responsible for the planning and preparation of a capable exascale ecosystem -- including software, applications, hardware, advanced system engineering, and early testbed platforms -- to support the nation's exascale computing imperative.

\section*{References}

\bibliography{library}

\begin{thebibliography}{53}
\providecommand{\natexlab}[1]{#1}
\providecommand{\url}[1]{\texttt{#1}}
\providecommand{\urlprefix}{URL }
\expandafter\ifx\csname urlstyle\endcsname\relax
  \providecommand{\doi}[1]{doi:\discretionary{}{}{}#1}\else
  \providecommand{\doi}[1]{doi:\discretionary{}{}{}\begingroup
  \urlstyle{rm}\url{#1}\endgroup}\fi
\providecommand{\bibinfo}[2]{#2}

\bibitem[{Day et~al.(2012)Day, Tachibana, Bell, Lijewski, Beckner, and
  Cheng}]{Day2012}
\bibinfo{author}{M.~Day}, \bibinfo{author}{S.~Tachibana},
  \bibinfo{author}{J.~Bell}, \bibinfo{author}{M.~Lijewski},
  \bibinfo{author}{V.~Beckner}, \bibinfo{author}{R.~K. Cheng},
  \bibinfo{title}{{A combined computational and experimental characterization
  of lean premixed turbulent low swirl laboratory flames}},
  \bibinfo{journal}{Combust. Flame} \bibinfo{volume}{159}~(\bibinfo{number}{1})
  (\bibinfo{year}{2012}) \bibinfo{pages}{275--290}, ISSN
  \bibinfo{issn}{00102180},
  \doi{\bibinfo{doi}{10.1016/j.combustflame.2011.06.016}},
  \urlprefix\url{http://linkinghub.elsevier.com/retrieve/pii/
  S0010218011001969}.

\bibitem[{Cook and Riley(1994)}]{Cook1994}
\bibinfo{author}{A.~W. Cook}, \bibinfo{author}{J.~J. Riley}, \bibinfo{title}{{A
  subgrid model for equilibrium chemistry in turbulent flows}},
  \bibinfo{journal}{Phys. Fluids} \bibinfo{volume}{6}~(\bibinfo{number}{8})
  (\bibinfo{year}{1994}) \bibinfo{pages}{2868--2870}, ISSN
  \bibinfo{issn}{1070-6631}, \doi{\bibinfo{doi}{10.1063/1.868111}},
  \urlprefix\url{http://aip.scitation.org/doi/10.1063/1.868111}.

\bibitem[{Jim{\'{e}}nez et~al.(1997)Jim{\'{e}}nez, Li{\~{n}}{\'{a}}n, Rogers,
  and Higuera}]{Jimenez1997}
\bibinfo{author}{J.~Jim{\'{e}}nez}, \bibinfo{author}{A.~Li{\~{n}}{\'{a}}n},
  \bibinfo{author}{M.~M. Rogers}, \bibinfo{author}{F.~J. Higuera},
  \bibinfo{title}{{A priori testing of subgrid models for chemically reacting
  non-premixed turbulent shear flows}}, \bibinfo{journal}{J. Fluid Mech.} ISSN
  \bibinfo{issn}{00221120}, \doi{\bibinfo{doi}{10.1017/S0022112097006733}}.

\bibitem[{Bradley et~al.(1998)Bradley, Gaskell, and Gu}]{Bradley1998}
\bibinfo{author}{D.~Bradley}, \bibinfo{author}{P.~Gaskell},
  \bibinfo{author}{X.~Gu}, \bibinfo{title}{{The mathematical modeling of
  liftoff and blowoff of turbulent non-premixed methane jet flames at high
  strain rates}}, \bibinfo{journal}{Symp. Combust.}
  \bibinfo{volume}{27}~(\bibinfo{number}{1}) (\bibinfo{year}{1998})
  \bibinfo{pages}{1199--1206}, ISSN \bibinfo{issn}{00820784},
  \doi{\bibinfo{doi}{10.1016/S0082-0784(98)80523-7}},
  \urlprefix\url{https://linkinghub.elsevier.com/retrieve/pii/
  S0082078498805237}.

\bibitem[{Bradley et~al.(2002)Bradley, Emerson, Gaskell, and Gu}]{Bradley2002}
\bibinfo{author}{D.~Bradley}, \bibinfo{author}{D.~Emerson},
  \bibinfo{author}{P.~Gaskell}, \bibinfo{author}{X.~Gu},
  \bibinfo{title}{{Mathematical modeling of turbulent non-premixed piloted-jet
  flames with local extinctions}}, \bibinfo{journal}{Proc. Combust. Inst.}
  \bibinfo{volume}{29}~(\bibinfo{number}{2}) (\bibinfo{year}{2002})
  \bibinfo{pages}{2155--2162}, ISSN \bibinfo{issn}{15407489},
  \doi{\bibinfo{doi}{10.1016/S1540-7489(02)80262-0}},
  \urlprefix\url{https://linkinghub.elsevier.com/retrieve/pii/
  S1540748902802620}.

\bibitem[{Ihme and Pitsch(2008{\natexlab{a}})}]{Ihme2008}
\bibinfo{author}{M.~Ihme}, \bibinfo{author}{H.~Pitsch},
  \bibinfo{title}{{Prediction of extinction and reignition in nonpremixed
  turbulent flames using a flamelet/progress variable model: 1. A priori study
  and presumed PDF closure}}, \bibinfo{journal}{Combust. Flame}
  \bibinfo{volume}{155}~(\bibinfo{number}{1-2})
  (\bibinfo{year}{2008}{\natexlab{a}}) \bibinfo{pages}{70--89}, ISSN
  \bibinfo{issn}{00102180},
  \doi{\bibinfo{doi}{10.1016/j.combustflame.2008.04.001}},
  \urlprefix\url{https://linkinghub.elsevier.com/retrieve/pii/
  S0010218008000904}.

\bibitem[{Ihme and Pitsch(2008{\natexlab{b}})}]{Ihme2008a}
\bibinfo{author}{M.~Ihme}, \bibinfo{author}{H.~Pitsch},
  \bibinfo{title}{{Prediction of extinction and reignition in nonpremixed
  turbulent flames using a flamelet/progress variable model: 2. Application in
  LES of Sandia flames D and E}}, \bibinfo{journal}{Combust. Flame}
  \bibinfo{volume}{155}~(\bibinfo{number}{1-2})
  (\bibinfo{year}{2008}{\natexlab{b}}) \bibinfo{pages}{90--107}, ISSN
  \bibinfo{issn}{00102180},
  \doi{\bibinfo{doi}{10.1016/j.combustflame.2008.04.015}},
  \urlprefix\url{https://linkinghub.elsevier.com/retrieve/pii/
  S0010218008001223}.

\bibitem[{Fern{\'{a}}ndez-Delgado et~al.(2014)Fern{\'{a}}ndez-Delgado,
  Cernadas, Barro, Amorim, and {Amorim
  Fern{\'{a}}ndez-Delgado}}]{Fernandez-Delgado2014}
\bibinfo{author}{M.~Fern{\'{a}}ndez-Delgado}, \bibinfo{author}{E.~Cernadas},
  \bibinfo{author}{S.~Barro}, \bibinfo{author}{D.~Amorim},
  \bibinfo{author}{D.~{Amorim Fern{\'{a}}ndez-Delgado}}, \bibinfo{title}{{Do we
  Need Hundreds of Classifiers to Solve Real World Classification Problems?}},
  \bibinfo{journal}{J. Mach. Learn. Res.} ISSN \bibinfo{issn}{1532-4435},
  \doi{\bibinfo{doi}{10.1016/j.csda.2008.10.033}}.

\bibitem[{Liaw and Wiener(2002)}]{Liaw2002}
\bibinfo{author}{A.~Liaw}, \bibinfo{author}{M.~Wiener},
  \bibinfo{title}{{Classification and Regression by randomForest}},
  \bibinfo{journal}{R news} \bibinfo{volume}{2}~(\bibinfo{number}{3})
  (\bibinfo{year}{2002}) \bibinfo{pages}{18--22}.

\bibitem[{Cybenko(1989)}]{Cybenko1989}
\bibinfo{author}{G.~Cybenko}, \bibinfo{title}{{Approximation by superpositions
  of a sigmoidal function}}, \bibinfo{journal}{Math. Control. Signals, Syst.}
  \bibinfo{volume}{2}~(\bibinfo{number}{4}) (\bibinfo{year}{1989})
  \bibinfo{pages}{303--314}, ISSN \bibinfo{issn}{0932-4194},
  \doi{\bibinfo{doi}{10.1007/BF02551274}},
  \urlprefix\url{http://link.springer.com/10.1007/BF02551274}.

\bibitem[{Hornik(1991)}]{Hornik1991}
\bibinfo{author}{K.~Hornik}, \bibinfo{title}{{Approximation capabilities of
  multilayer feedforward networks}}, \bibinfo{journal}{Neural Networks}
  \bibinfo{volume}{4}~(\bibinfo{number}{2}) (\bibinfo{year}{1991})
  \bibinfo{pages}{251--257}, ISSN \bibinfo{issn}{08936080},
  \doi{\bibinfo{doi}{10.1016/0893-6080(91)90009-T}},
  \urlprefix\url{http://linkinghub.elsevier.com/retrieve/pii/
  089360809190009T}.

\bibitem[{Goodfellow et~al.(2016)Goodfellow, Bengio, and
  Courville}]{Goodfellow2016}
\bibinfo{author}{I.~Goodfellow}, \bibinfo{author}{Y.~Bengio},
  \bibinfo{author}{A.~Courville}, \bibinfo{title}{{Deep Learning}},
  \bibinfo{publisher}{MIT Press},
  \urlprefix\url{http://www.deeplearningbook.org}, \bibinfo{year}{2016}.

\bibitem[{Veynante and Vervisch(2002)}]{Veynante2002}
\bibinfo{author}{D.~Veynante}, \bibinfo{author}{L.~Vervisch},
  \bibinfo{title}{{Turbulent combustion modeling}}, \bibinfo{journal}{Prog.
  Energy Combust. Sci.} \bibinfo{volume}{28}~(\bibinfo{number}{3})
  (\bibinfo{year}{2002}) \bibinfo{pages}{193--266}, ISSN
  \bibinfo{issn}{03601285}, \doi{\bibinfo{doi}{10.1016/S0360-1285(01)00017-X}},
  \urlprefix\url{http://linkinghub.elsevier.com/retrieve/pii/
  S036012850100017X}.

\bibitem[{Pitsch(2006)}]{Pitsch2006a}
\bibinfo{author}{H.~Pitsch}, \bibinfo{title}{{LARGE-EDDY SIMULATION OF
  TURBULENT COMBUSTION}}, \bibinfo{journal}{Annu. Rev. Fluid Mech.}
  \bibinfo{volume}{38}~(\bibinfo{number}{1}) (\bibinfo{year}{2006})
  \bibinfo{pages}{453--482}, ISSN \bibinfo{issn}{0066-4189},
  \doi{\bibinfo{doi}{10.1146/annurev.fluid.38.050304.092133}},
  \urlprefix\url{http://www.annualreviews.org/doi/10.1146/
  annurev.fluid.38.050304.092133}.

\bibitem[{van Oijen(2002)}]{VanOijen2002}
\bibinfo{author}{J.~van Oijen}, \bibinfo{title}{{Flamelet-Generated Manifolds:
  Development and Application to Premixed Laminar Flames}}, Ph.D. thesis,
  \bibinfo{school}{Eindhoven University of Technology},
  \bibinfo{address}{Eindhoven}, \bibinfo{year}{2002}.

\bibitem[{Gicquel et~al.(2000)Gicquel, Darabiha, and
  Th{\'{e}}venin}]{Gicquel2000}
\bibinfo{author}{O.~Gicquel}, \bibinfo{author}{N.~Darabiha},
  \bibinfo{author}{D.~Th{\'{e}}venin}, \bibinfo{title}{{Liminar premixed
  hydrogen/air counterflow flame simulations using flame prolongation of ILDM
  with differential diffusion}}, \bibinfo{journal}{Proc. Combust. Inst.}
  \bibinfo{volume}{28}~(\bibinfo{number}{2}) (\bibinfo{year}{2000})
  \bibinfo{pages}{1901--1908}, ISSN \bibinfo{issn}{15407489},
  \doi{\bibinfo{doi}{10.1016/S0082-0784(00)80594-9}},
  \urlprefix\url{http://linkinghub.elsevier.com/retrieve/pii/
  S0082078400805949}.

\bibitem[{Klimenko and Bilger(1999)}]{Klimenko1999}
\bibinfo{author}{A.~Y. Klimenko}, \bibinfo{author}{R.~W. Bilger},
  \bibinfo{title}{{Conditional moment closure for turbulent combustion}},
  \bibinfo{journal}{Prog. Energy Combust. Sci.} ISSN \bibinfo{issn}{03601285},
  \doi{\bibinfo{doi}{10.1016/S0360-1285(99)00006-4}}.

\bibitem[{Jin et~al.(2008)Jin, Grout, and Bushe}]{JinGB08}
\bibinfo{author}{B.~Jin}, \bibinfo{author}{R.~Grout}, \bibinfo{author}{W.~K.
  Bushe}, \bibinfo{title}{Conditional Source-Term Estimation as a Method for
  Chemical Closure in Premixed Turbulent Reacting Flow},
  \bibinfo{journal}{Flow, Turbulence and Combustion}
  \bibinfo{volume}{81}~(\bibinfo{number}{4}) (\bibinfo{year}{2008})
  \bibinfo{pages}{563--582}, ISSN \bibinfo{issn}{1573-1987},
  \doi{\bibinfo{doi}{10.1007/s10494-008-9148-0}},
  \urlprefix\url{https://doi.org/10.1007/s10494-008-9148-0}.

\bibitem[{Fox(2003)}]{Fox2003}
\bibinfo{author}{R.~O. Fox}, \bibinfo{title}{{Computational Models for
  Turbulent Reacting Flows}}, \bibinfo{publisher}{Cambridge Univ. Press},
  \bibinfo{address}{Cambridge, UK}, \bibinfo{year}{2003}.

\bibitem[{Grout et~al.(2009)Grout, Swaminathan, and Cant}]{Grout2009}
\bibinfo{author}{R.~W. Grout}, \bibinfo{author}{N.~Swaminathan},
  \bibinfo{author}{R.~S. Cant}, \bibinfo{title}{{Effects of compositional
  fluctuations on premixed flames}}, \bibinfo{journal}{Combust. Theory Model.}
  \bibinfo{volume}{13}~(\bibinfo{number}{5}) (\bibinfo{year}{2009})
  \bibinfo{pages}{823--852}, ISSN \bibinfo{issn}{1364-7830},
  \doi{\bibinfo{doi}{10.1080/13647830903160291}},
  \urlprefix\url{http://www.tandfonline.com/doi/abs/10.1080/
  13647830903160291}.

\bibitem[{Isaac et~al.(2014)Isaac, Coussement, Gicquel, Smith, and
  Parente}]{Isaac2014}
\bibinfo{author}{B.~J. Isaac}, \bibinfo{author}{A.~Coussement},
  \bibinfo{author}{O.~Gicquel}, \bibinfo{author}{P.~J. Smith},
  \bibinfo{author}{A.~Parente}, \bibinfo{title}{{Reduced-order PCA models for
  chemical reacting flows}}, \bibinfo{journal}{Combust. Flame}
  \bibinfo{volume}{161}~(\bibinfo{number}{11}) (\bibinfo{year}{2014})
  \bibinfo{pages}{2785--2800}, ISSN \bibinfo{issn}{00102180},
  \doi{\bibinfo{doi}{10.1016/j.combustflame.2014.05.011}},
  \urlprefix\url{https://linkinghub.elsevier.com/retrieve/pii/
  S0010218014001412}.

\bibitem[{Linse et~al.(2014)Linse, Kleemann, and Hasse}]{Linse2014}
\bibinfo{author}{D.~Linse}, \bibinfo{author}{A.~Kleemann},
  \bibinfo{author}{C.~Hasse}, \bibinfo{title}{{Probability density function
  approach coupled with detailed chemical kinetics for the prediction of knock
  in turbocharged direct injection spark ignition engines}},
  \bibinfo{journal}{Combust. Flame} \bibinfo{volume}{161}~(\bibinfo{number}{4})
  (\bibinfo{year}{2014}) \bibinfo{pages}{997--1014}, ISSN
  \bibinfo{issn}{00102180},
  \doi{\bibinfo{doi}{10.1016/j.combustflame.2013.10.025}},
  \urlprefix\url{https://linkinghub.elsevier.com/retrieve/pii/
  S0010218013003994}.

\bibitem[{Perry and Mueller(2018)}]{Perry2018}
\bibinfo{author}{B.~A. Perry}, \bibinfo{author}{M.~E. Mueller},
  \bibinfo{title}{{Joint probability density function models for multiscalar
  turbulent mixing}}, \bibinfo{journal}{Combust. Flame} \bibinfo{volume}{193}
  (\bibinfo{year}{2018}) \bibinfo{pages}{344--362}, ISSN
  \bibinfo{issn}{00102180},
  \doi{\bibinfo{doi}{10.1016/j.combustflame.2018.03.039}},
  \urlprefix\url{https://doi.org/10.1016/j.combustflame.2018.03.039 https://
  linkinghub.elsevier.com/retrieve/pii/S0010218018301512}.

\bibitem[{Cheng et~al.(2000)Cheng, Yegian, Miyasato, Samuelsen, Benson,
  Pellizzari, and Loftus}]{Cheng2000}
\bibinfo{author}{R.~Cheng}, \bibinfo{author}{D.~Yegian},
  \bibinfo{author}{M.~Miyasato}, \bibinfo{author}{G.~Samuelsen},
  \bibinfo{author}{C.~Benson}, \bibinfo{author}{R.~Pellizzari},
  \bibinfo{author}{P.~Loftus}, \bibinfo{title}{{Scaling and development of
  low-swirl burners for low-emission furnaces and boilers}},
  \bibinfo{journal}{Proc. Combust. Inst.} ISSN \bibinfo{issn}{15407489},
  \doi{\bibinfo{doi}{10.1016/S0082-0784(00)80344-6}}.

\bibitem[{Day and Bell(2000)}]{Day2000}
\bibinfo{author}{M.~S. Day}, \bibinfo{author}{J.~B. Bell},
  \bibinfo{title}{{Numerical simulation of laminar reacting flows with complex
  chemistry}}, \bibinfo{journal}{Combust. Theory Model.}
  \bibinfo{volume}{4}~(\bibinfo{number}{4}) (\bibinfo{year}{2000})
  \bibinfo{pages}{535--556}, ISSN \bibinfo{issn}{1364-7830},
  \doi{\bibinfo{doi}{10.1088/1364-7830/4/4/309}},
  \urlprefix\url{http://www.tandfonline.com/doi/abs/10.1088/1364-7830/4/4/
  309}.

\bibitem[{Kazakov and Frenklach(1994)}]{Kazakov1994}
\bibinfo{author}{A.~Kazakov}, \bibinfo{author}{M.~Frenklach},
  \bibinfo{title}{{Reduced Reaction Sets based on GRI-Mech 1.2}},
  \urlprefix\url{http://www.me.berkeley.edu/drm/}, \bibinfo{year}{1994}.

\bibitem[{Pedregosa et~al.(2011)Pedregosa, Varoquaux, Gramfort, Michel,
  Thirion, Grisel, Blondel, Prettenhofer, Weiss, Dubourg, Vanderplas, Passos,
  Cournapeau, Brucher, Perrot, and Duchesnay}]{Pedregosa2011}
\bibinfo{author}{F.~Pedregosa}, \bibinfo{author}{G.~Varoquaux},
  \bibinfo{author}{A.~Gramfort}, \bibinfo{author}{V.~Michel},
  \bibinfo{author}{B.~Thirion}, \bibinfo{author}{O.~Grisel},
  \bibinfo{author}{M.~Blondel}, \bibinfo{author}{P.~Prettenhofer},
  \bibinfo{author}{R.~Weiss}, \bibinfo{author}{V.~Dubourg},
  \bibinfo{author}{J.~Vanderplas}, \bibinfo{author}{A.~Passos},
  \bibinfo{author}{D.~Cournapeau}, \bibinfo{author}{M.~Brucher},
  \bibinfo{author}{M.~Perrot}, \bibinfo{author}{E.~Duchesnay},
  \bibinfo{title}{{Scikit-learn: Machine Learning in Python}},
  \bibinfo{journal}{J. Mach. Learn. Res.} ISSN \bibinfo{issn}{1271-6669},
  \doi{\bibinfo{doi}{10.1007/s13398-014-0173-7.2}}.

\bibitem[{Breiman(2001)}]{Breiman2001}
\bibinfo{author}{L.~Breiman}, \bibinfo{title}{{Random forests}},
  \bibinfo{journal}{Mach. Learn.} ISSN \bibinfo{issn}{08856125},
  \doi{\bibinfo{doi}{10.1023/A:1010933404324}}.

\bibitem[{Goodfellow et~al.(2014)Goodfellow, Pouget-Abadie, Mirza, Xu,
  Warde-Farley, Ozair, Courville, and Bengio}]{Goodfellow2014}
\bibinfo{author}{I.~Goodfellow}, \bibinfo{author}{J.~Pouget-Abadie},
  \bibinfo{author}{M.~Mirza}, \bibinfo{author}{B.~Xu},
  \bibinfo{author}{D.~Warde-Farley}, \bibinfo{author}{S.~Ozair},
  \bibinfo{author}{A.~Courville}, \bibinfo{author}{Y.~Bengio},
  \bibinfo{title}{{Generative Adversarial Nets}}, \bibinfo{journal}{Adv. Neural
  Inf. Process. Syst. 27}  (\bibinfo{year}{2014})
  \bibinfo{pages}{2672--2680}ISSN \bibinfo{issn}{10495258},
  \doi{\bibinfo{doi}{10.1017/CBO9781139058452}},
  \urlprefix\url{http://papers.nips.cc/paper/5423-generative-adversarial-
  nets.pdf}.

\bibitem[{Burger et~al.(2012)Burger, Schuler, and Harmeling}]{Burger2012}
\bibinfo{author}{H.~C. Burger}, \bibinfo{author}{C.~J. Schuler},
  \bibinfo{author}{S.~Harmeling}, \bibinfo{title}{{Image denoising: Can plain
  neural networks compete with BM3D?}}, in: \bibinfo{booktitle}{Proc. IEEE
  Comput. Soc. Conf. Comput. Vis. Pattern Recognit.}, ISBN
  \bibinfo{isbn}{9781467312264}, ISSN \bibinfo{issn}{10636919},
  \bibinfo{pages}{2392--2399}, \doi{\bibinfo{doi}{10.1109/CVPR.2012.6247952}},
  \bibinfo{year}{2012}.

\bibitem[{Dosovitskiy et~al.(2015)Dosovitskiy, Springenberg, and
  Brox}]{Dosovitskiy2015}
\bibinfo{author}{A.~Dosovitskiy}, \bibinfo{author}{J.~T. Springenberg},
  \bibinfo{author}{T.~Brox}, \bibinfo{title}{{Learning to generate chairs with
  convolutional neural networks}}, in: \bibinfo{booktitle}{Proc. IEEE Comput.
  Soc. Conf. Comput. Vis. Pattern Recognit.}, vol.
  \bibinfo{volume}{07-12-June}, ISBN \bibinfo{isbn}{9781467369640}, ISSN
  \bibinfo{issn}{10636919}, \bibinfo{pages}{1538--1546},
  \doi{\bibinfo{doi}{10.1109/CVPR.2015.7298761}}, \bibinfo{year}{2015}.

\bibitem[{Lefkimmiatis(2016)}]{Lefkimmiatis2016}
\bibinfo{author}{S.~Lefkimmiatis}, \bibinfo{title}{{Non-Local Color Image
  Denoising with Convolutional Neural Networks}}, \bibinfo{journal}{Proc. IEEE
  Conf. Comput. Vis. Pattern Recognit.}  (\bibinfo{year}{2016})
  \bibinfo{pages}{5882--5891}\doi{\bibinfo{doi}{10.1109/CVPR.2017.623}},
  \urlprefix\url{http://arxiv.org/abs/1611.06757}.

\bibitem[{Ledig et~al.(2016)Ledig, Theis, Huszar, Caballero, Cunningham,
  Acosta, Aitken, Tejani, Totz, Wang, and Shi}]{Ledig2017}
\bibinfo{author}{C.~Ledig}, \bibinfo{author}{L.~Theis},
  \bibinfo{author}{F.~Huszar}, \bibinfo{author}{J.~Caballero},
  \bibinfo{author}{A.~Cunningham}, \bibinfo{author}{A.~Acosta},
  \bibinfo{author}{A.~Aitken}, \bibinfo{author}{A.~Tejani},
  \bibinfo{author}{J.~Totz}, \bibinfo{author}{Z.~Wang},
  \bibinfo{author}{W.~Shi}, \bibinfo{title}{{Photo-Realistic Single Image
  Super-Resolution Using a Generative Adversarial Network}},
  \bibinfo{journal}{Conf. Comput. Vis. Pattern Recognit.}
  (\bibinfo{year}{2016}) \bibinfo{pages}{1--14}ISSN \bibinfo{issn}{0018-5043},
  \doi{\bibinfo{doi}{10.1109/CVPR.2017.19}},
  \urlprefix\url{http://arxiv.org/abs/1609.04802}.

\bibitem[{Tai et~al.(2017)Tai, Yang, and Liu}]{Tai2017}
\bibinfo{author}{Y.~Tai}, \bibinfo{author}{J.~Yang}, \bibinfo{author}{X.~Liu},
  \bibinfo{title}{{Image super-resolution via deep recursive residual
  network}}, in: \bibinfo{booktitle}{Proc. - 30th IEEE Conf. Comput. Vis.
  Pattern Recognition, CVPR 2017}, vol. \bibinfo{volume}{2017-Janua}, ISBN
  \bibinfo{isbn}{9781538604571}, \bibinfo{pages}{2790--2798},
  \doi{\bibinfo{doi}{10.1109/CVPR.2017.298}}, \bibinfo{year}{2017}.

\bibitem[{Lai et~al.(2017)Lai, Huang, Ahuja, and Yang}]{Lai2017}
\bibinfo{author}{W.~S. Lai}, \bibinfo{author}{J.~B. Huang},
  \bibinfo{author}{N.~Ahuja}, \bibinfo{author}{M.~H. Yang},
  \bibinfo{title}{{Deep laplacian pyramid networks for fast and accurate
  super-resolution}}, in: \bibinfo{booktitle}{Proc. - 30th IEEE Conf. Comput.
  Vis. Pattern Recognition, CVPR 2017}, vol. \bibinfo{volume}{2017-Janua}, ISBN
  \bibinfo{isbn}{9781538604571}, ISSN \bibinfo{issn}{1063-6919},
  \bibinfo{pages}{5835--5843}, \doi{\bibinfo{doi}{10.1109/CVPR.2017.618}},
  \bibinfo{year}{2017}.

\bibitem[{Graves(2013)}]{Graves2013}
\bibinfo{author}{A.~Graves}, \bibinfo{title}{{Generating sequences with
  recurrent neural networks. preprint}}, \bibinfo{journal}{arXiv:1308.0850}
  ISSN \bibinfo{issn}{18792782}, \doi{\bibinfo{doi}{10.1145/2661829.2661935}}.

\bibitem[{Wu et~al.(2016)Wu, Schuster, Chen, Le, Norouzi, Macherey, Krikun,
  Cao, Gao, Macherey, Klingner, Shah, Johnson, Liu, Kaiser, Gouws, Kato, Kudo,
  Kazawa, Stevens, Kurian, Patil, Wang, Young, Smith, Riesa, Rudnick, Vinyals,
  Corrado, Hughes, and Dean}]{Wu2016}
\bibinfo{author}{Y.~Wu}, \bibinfo{author}{M.~Schuster},
  \bibinfo{author}{Z.~Chen}, \bibinfo{author}{Q.~V. Le},
  \bibinfo{author}{M.~Norouzi}, \bibinfo{author}{W.~Macherey},
  \bibinfo{author}{M.~Krikun}, \bibinfo{author}{Y.~Cao},
  \bibinfo{author}{Q.~Gao}, \bibinfo{author}{K.~Macherey},
  \bibinfo{author}{J.~Klingner}, \bibinfo{author}{A.~Shah},
  \bibinfo{author}{M.~Johnson}, \bibinfo{author}{X.~Liu},
  \bibinfo{author}{{\L}.~Kaiser}, \bibinfo{author}{S.~Gouws},
  \bibinfo{author}{Y.~Kato}, \bibinfo{author}{T.~Kudo},
  \bibinfo{author}{H.~Kazawa}, \bibinfo{author}{K.~Stevens},
  \bibinfo{author}{G.~Kurian}, \bibinfo{author}{N.~Patil},
  \bibinfo{author}{W.~Wang}, \bibinfo{author}{C.~Young},
  \bibinfo{author}{J.~Smith}, \bibinfo{author}{J.~Riesa},
  \bibinfo{author}{A.~Rudnick}, \bibinfo{author}{O.~Vinyals},
  \bibinfo{author}{G.~Corrado}, \bibinfo{author}{M.~Hughes},
  \bibinfo{author}{J.~Dean}, \bibinfo{title}{{Google's Neural Machine
  Translation System: Bridging the Gap between Human and Machine Translation}},
  \bibinfo{journal}{ArXiv e-prints} ISSN \bibinfo{issn}{1471003X},
  \doi{\bibinfo{doi}{10.1038/nrn2258}}.

\bibitem[{Kwon(2017)}]{Kwon2017}
\bibinfo{author}{W.~Kwon}, \bibinfo{title}{{Attention Is All You Need}},
  \bibinfo{journal}{arXiv1706.03762 [cs]} .

\bibitem[{Silver et~al.(2017)Silver, Schrittwieser, Simonyan, Antonoglou,
  Huang, Guez, Hubert, Baker, Lai, Bolton, Chen, Lillicrap, Hui, Sifre, van~den
  Driessche, Graepel, and Hassabis}]{Silver2017}
\bibinfo{author}{D.~Silver}, \bibinfo{author}{J.~Schrittwieser},
  \bibinfo{author}{K.~Simonyan}, \bibinfo{author}{I.~Antonoglou},
  \bibinfo{author}{A.~Huang}, \bibinfo{author}{A.~Guez},
  \bibinfo{author}{T.~Hubert}, \bibinfo{author}{L.~Baker},
  \bibinfo{author}{M.~Lai}, \bibinfo{author}{A.~Bolton},
  \bibinfo{author}{Y.~Chen}, \bibinfo{author}{T.~Lillicrap},
  \bibinfo{author}{F.~Hui}, \bibinfo{author}{L.~Sifre},
  \bibinfo{author}{G.~van~den Driessche}, \bibinfo{author}{T.~Graepel},
  \bibinfo{author}{D.~Hassabis}, \bibinfo{title}{{Mastering the game of Go
  without human knowledge}}, \bibinfo{journal}{Nature}
  \bibinfo{volume}{550}~(\bibinfo{number}{7676}) (\bibinfo{year}{2017})
  \bibinfo{pages}{354--359}, ISSN \bibinfo{issn}{0028-0836},
  \doi{\bibinfo{doi}{10.1038/nature24270}},
  \urlprefix\url{http://www.nature.com/doifinder/10.1038/nature24270}.

\bibitem[{Lecun et~al.(2015)Lecun, Bengio, and Hinton}]{Lecun2015}
\bibinfo{author}{Y.~Lecun}, \bibinfo{author}{Y.~Bengio},
  \bibinfo{author}{G.~Hinton}, \bibinfo{title}{{Deep learning}},
  \bibinfo{journal}{Nature} \bibinfo{volume}{521}~(\bibinfo{number}{7553})
  (\bibinfo{year}{2015}) \bibinfo{pages}{436--444}, ISSN
  \bibinfo{issn}{14764687}, \doi{\bibinfo{doi}{10.1038/nature14539}}.

\bibitem[{Schmidhuber(2015)}]{Schmidhuber2015}
\bibinfo{author}{J.~Schmidhuber}, \bibinfo{title}{{Deep Learning in neural
  networks: An overview}}, \bibinfo{journal}{Neural Networks}
  \bibinfo{volume}{61} (\bibinfo{year}{2015}) \bibinfo{pages}{85--117}, ISSN
  \bibinfo{issn}{18792782}, \doi{\bibinfo{doi}{10.1016/j.neunet.2014.09.003}},
  \urlprefix\url{http://dx.doi.org/10.1016/j.neunet.2014.09.003}.

\bibitem[{Prieto et~al.(2016)Prieto, Prieto, Ortigosa, Ros, Pelayo, Ortega, and
  Rojas}]{Prieto2016}
\bibinfo{author}{A.~Prieto}, \bibinfo{author}{B.~Prieto},
  \bibinfo{author}{E.~M. Ortigosa}, \bibinfo{author}{E.~Ros},
  \bibinfo{author}{F.~Pelayo}, \bibinfo{author}{J.~Ortega},
  \bibinfo{author}{I.~Rojas}, \bibinfo{title}{{Neural networks: An overview of
  early research, current frameworks and new challenges}},
  \bibinfo{journal}{Neurocomputing} ISSN \bibinfo{issn}{18728286},
  \doi{\bibinfo{doi}{10.1016/j.neucom.2016.06.014}}.

\bibitem[{Liu et~al.(2017)Liu, Wang, Liu, Zeng, Liu, and Alsaadi}]{Liu2017}
\bibinfo{author}{W.~Liu}, \bibinfo{author}{Z.~Wang}, \bibinfo{author}{X.~Liu},
  \bibinfo{author}{N.~Zeng}, \bibinfo{author}{Y.~Liu}, \bibinfo{author}{F.~E.
  Alsaadi}, \bibinfo{title}{{A survey of deep neural network architectures and
  their applications}}, \bibinfo{journal}{Neurocomputing} ISSN
  \bibinfo{issn}{18728286}, \doi{\bibinfo{doi}{10.1016/j.neucom.2016.12.038}}.

\bibitem[{Ioffe and Szegedy(2015)}]{Ioffe2015}
\bibinfo{author}{S.~Ioffe}, \bibinfo{author}{C.~Szegedy},
  \bibinfo{title}{{Batch Normalization: Accelerating Deep Network Training by
  Reducing Internal Covariate Shift}}
  \urlprefix\url{http://arxiv.org/abs/1502.03167}.

\bibitem[{Kingma and Ba(2014)}]{Kingma2014}
\bibinfo{author}{D.~P. Kingma}, \bibinfo{author}{J.~Ba}, \bibinfo{title}{{Adam:
  A Method for Stochastic Optimization}}
  \urlprefix\url{http://arxiv.org/abs/1412.6980}.

\bibitem[{Paszke et~al.(2017)Paszke, Gross, Chintala, Chanan, Yang, DeVito,
  Lin, Desmaison, Antiga, and Lerer}]{Paszke2017}
\bibinfo{author}{A.~Paszke}, \bibinfo{author}{S.~Gross},
  \bibinfo{author}{S.~Chintala}, \bibinfo{author}{G.~Chanan},
  \bibinfo{author}{E.~Yang}, \bibinfo{author}{Z.~DeVito},
  \bibinfo{author}{Z.~Lin}, \bibinfo{author}{A.~Desmaison},
  \bibinfo{author}{L.~Antiga}, \bibinfo{author}{A.~Lerer},
  \bibinfo{title}{{Automatic differentiation in PyTorch}}, in:
  \bibinfo{booktitle}{NIPS-W}, \bibinfo{year}{2017}.

\bibitem[{Kingma and Welling(2013)}]{Kingma2013}
\bibinfo{author}{D.~P. Kingma}, \bibinfo{author}{M.~Welling},
  \bibinfo{title}{{Auto-Encoding Variational Bayes}}
  \urlprefix\url{http://arxiv.org/abs/1312.6114}.

\bibitem[{Rezende et~al.(2014)Rezende, Mohamed, and Wierstra}]{Rezende2014}
\bibinfo{author}{D.~J. Rezende}, \bibinfo{author}{S.~Mohamed},
  \bibinfo{author}{D.~Wierstra}, \bibinfo{title}{{Stochastic Backpropagation
  and Approximate Inference in Deep Generative Models}}
  \urlprefix\url{http://arxiv.org/abs/1401.4082}.

\bibitem[{Chen et~al.(2016)Chen, Duan, Houthooft, Schulman, Sutskever, and
  Abbeel}]{Chen2016a}
\bibinfo{author}{X.~Chen}, \bibinfo{author}{Y.~Duan},
  \bibinfo{author}{R.~Houthooft}, \bibinfo{author}{J.~Schulman},
  \bibinfo{author}{I.~Sutskever}, \bibinfo{author}{P.~Abbeel},
  \bibinfo{title}{{InfoGAN: Interpretable Representation Learning by
  Information Maximizing Generative Adversarial Nets}}
  \urlprefix\url{http://arxiv.org/abs/1606.03657}.

\bibitem[{Endres and Schindelin(2003)}]{Endres2003}
\bibinfo{author}{D.~M. Endres}, \bibinfo{author}{J.~E. Schindelin},
  \bibinfo{title}{{A new metric for probability distributions}},
  \doi{\bibinfo{doi}{10.1109/TIT.2003.813506}}, \bibinfo{year}{2003}.

\bibitem[{{\"{O}}sterreicher and Vajda(2003)}]{Osterreicher2003}
\bibinfo{author}{F.~{\"{O}}sterreicher}, \bibinfo{author}{I.~Vajda},
  \bibinfo{title}{{A new class of metric divergences on probability spaces and
  its applicability in statistics}}, \bibinfo{journal}{Ann. Inst. Stat. Math.}
  ISSN \bibinfo{issn}{00203157}, \doi{\bibinfo{doi}{10.1007/BF02517812}}.

\bibitem[{Kullback(1987)}]{Kullback1987}
\bibinfo{author}{S.~Kullback}, \bibinfo{title}{{Letters to the Editor}},
  \bibinfo{journal}{Am. Stat.} \bibinfo{volume}{41}~(\bibinfo{number}{4})
  (\bibinfo{year}{1987}) \bibinfo{pages}{338--341}, ISSN
  \bibinfo{issn}{0003-1305},
  \doi{\bibinfo{doi}{10.1080/00031305.1987.10475510}},
  \urlprefix\url{http://www.tandfonline.com/doi/abs/10.1080/
  00031305.1987.10475510}.

\bibitem[{Jones et~al.(01  )Jones, Oliphant, Peterson et~al.}]{Jones2001}
\bibinfo{author}{E.~Jones}, \bibinfo{author}{T.~Oliphant},
  \bibinfo{author}{P.~Peterson}, et~al., \bibinfo{title}{{SciPy}: Open source
  scientific tools for {Python}}, \urlprefix\url{http://www.scipy.org/},
  \bibinfo{note}{[Online; accessed 12/10/2018]}, \bibinfo{year}{2001--}.

\end{thebibliography}

\appendix
\section{Summary of model parameters}\label{app:hp_summary}
The random forests model parameters are: number of estimators = 100,
max.\ depth = 30, criterion is MSE, min.\ samples required to split
node = 2, min.\ samples required for a leaf node = 1, max.\ number of
features for best split = all, max.\ number of leaf nodes is
unconstrained, bootstrapping when building trees is on, use out-of-bag
samples to estimate the $R^2$ is off.

The \gls{dnn} architecture is:
\begin{enumerate}
\item Linear layer (input features=4, output features=256, with bias)
\item LeakyReLU activation function with ($\alpha=0.01$)
\item BatchNorm1d (input and output features = 256, $\epsilon = 10^{-5}$, decay of 0.1)
\item Linear layer (input features=256, output features=512, with bias)
\item LeakyReLU activation function with ($\alpha=0.01$)
\item BatchNorm1d (input and output features = 512, $\epsilon = 10^{-5}$, decay of 0.1)
\item Linear layer (input features=512, output features=2048, with bias)
\item Softmax activation function
\end{enumerate}
Additional \gls{dnn} hyperparameters are: the learning rate
($10^{-4}$), the number of epochs (500), the batch size (64).

The \gls{cvae} encoder architecture and latent is:
\begin{enumerate}
\item Linear layer (input features=2052, output features=512, with bias)
\item ReLU activation function
\item Linear layer (input features=512, output features=256, with bias)
\item ReLU activation function
\item Latent space linear layer of means and variances (input features=256, output features=10, with bias)
\end{enumerate}
The \gls{cvae} decoder architecture is:
\begin{enumerate}
\item Linear layer (input features=14, output features=256, with bias)
\item ReLU activation function
\item Linear layer (input features=256, output features=512, with bias)
\item ReLU activation function
\item Linear layer (input features=512, output features=2048, with bias)
\item Softmax activation function
\end{enumerate}
Additional \gls{cvae} hyperparameters are: the learning rate
($10^{-3}$), the number of epochs (500), the batch size (64).
\end{document}